\documentclass{aastex63}

\usepackage{tabularx}  \usepackage{booktabs}  \usepackage{amsmath}   \usepackage{graphicx,subfigure}

\begin{document}

\title{The Yin-Yang Magnetic Flux Eruption (Yin-Yang-MFE) Code: A Global Corona Magnetohydrodynamic Code with the Yin-Yang grid}

\correspondingauthor{Yuhong Fan}
\email{yfan@ucar.edu}

\author{Hongyang Luo}
\affiliation{{Department of Earth and Planetary Sciences, the University of Hong Kong, Pokfulam, Hong Kong SAR}}
\affiliation{High Altitude Observatory, National Center for Atmospheric Research, United States of America}

\author{Yuhong Fan}
\affiliation{High Altitude Observatory, National Center for Atmospheric Research, United States of America}

\begin{abstract}

We describe the numerical algorithms of a global magnetohydrodynamic (MHD) code utilizing the Yin-Yang grid, called the Yin-Yang Magnetic Flux Eruption (Yin-Yang-MFE) code, suitable for modeling the large-scale dynamical processes of the solar corona and the solar wind. It is a single-fluid MHD code taking into account the non-adiabatic effects of the solar corona, including the electron heat conduction, optically thin radiative cooling, and empirical coronal heating.  We describe the numerical algorithms used to solve the set of MHD equations (with the semi-relativistic correction, or the Boris correction) in each of the partial spherical shell Yin Yang domains, and the method for updating the boundary conditions in the ghost-zones of the two overlapping domains with the code parallelized with the message passing interface (MPI).  We validate the code performance with a set of standard test problems, and finally present a solar wind solution with a dipolar magnetic flux distribution at the solar surface, representative of solar minimum configuration.

\end{abstract}


\section{Introduction} \label{sec:intro}

Numerical magnetohydrodynamic (MHD) simulations have become indispensable for studying astrophysical plasma flows, and they serve as a workhorse for investigating phenomena across a wide range of environments. Over the past few decades, many sophisticated MHD codes have been developed which have wide-ranging applicabilities, several examples include the ZEUS code \citep{stone1992zeus}, the Athena framework \citep{stone2008athena,stone2020athena++}, the PLUTO code \citep{mignone2007pluto}, the BATS-R-US code \citep{Toth:2012}, the Stagger code \citep{stein2024stagger}, the Pencil code \citep{dobler2006magnetic}, the MURaM code \citep{rempel2016extension}, the MPI-AMRVAC code \citep{keppens2012parallel}, and the R2D2 code \citep{hotta2022generation}. The Magnetic Flux Eruption (MFE) code as a relatively simple MHD code has been used for more than a decade as a specialized and robust tool to investigate the evolution of magnetic field in the Sun’s interior and the corona. In particular, it has been applied to model the subsurface rise of twisted active-region flux tubes, their emergence into the solar atmosphere, and the initiation and evolution of coronal mass ejections (CMEs) \citep[e.g][]{fan2009emergence,Chatterjee:Fan:2013,fan2017mhd,fan2018mhd,fan2022improved}. Recently it has been incorporated with boundary data-driven capability using the observed photospheric vetcor magnetograms to model realistic CME events \citep{Fan:etal:2024}. These studies demonstrate the code’s ability to capture the complex dynamics of magnetic flux ropes and eruptions in the solar context, providing valuable insights into the formation and structures of CME precursors and the mechanisms for their eruption.

Building on this foundation and with the purpose of open science, we present in this paper an enhanced version of the MFE code, referred to here as the Yin-Yang MFE code, which incorporates several important improvements to its numerical algorithms and grid structure. In developing the Yin-Yang MFE code, we placed special emphasis on achieving robust and accurate performance under low plasma-$\beta$ conditions (where magnetic pressure dominates gas pressure). Such low-$\beta$ environments, characteristic of the solar corona, pose well-known numerical challenges for MHD simulations. For example, explicit time-stepping can be severely limited by extremely high Alfv{\'e}n speeds, numerical instabilities (such as negative pressure values) may arise if the scheme is not sufficiently robust, and any failure to strictly maintain the divergence-free condition ($\nabla \cdot \mathbf{B} = 0$) can result in unphysical field-aligned Lorentz forces, particularly when a conservative formulation is used \citep{brackbill1980effect}. To mitigate these issues, the Yin-Yang MFE code employs several numerical strategies. We have implemented a high-order spatial reconstruction scheme and a numerical diffusive flux based on monotonicity constraints, which minimizes numerical diffusion that is crucial for resolving fine-scale structures without excessive smoothing. We also use a constrained transport (CT) approach with an improved evaluation of the electric field that minimizes magnetic diffusion at cell edges, ensuring that the magnetic field remains divergence-free to machine precision throughout the simulation.

Another key development in the Yin-Yang MFE code is the introduction of an overset Yin-Yang grid for the computational domain. The Yin-Yang grid consists of two perpendicular overlapping grid patches that together cover the full spherical surface without singularities. By adopting this grid system, the code can now perform full-sphere simulations without encountering the coordinate singularity and severe time-step constraints that affect traditional spherical grids near the poles. This capability marks a major extension of the original MFE code, enabling global simulations of the solar interior and corona in a single continuous domain. In particular, it allows us to model phenomena that involve interactions between high-latitude and low-latitude regions (or polar connections) with greater fidelity than was previously possible.

The remainder of this paper is organized as follows. In Section~\ref{sec:equations}, we describe the system of MHD equations solved by the Yin-Yang MFE code. Section~\ref{sec:algorithm} details the numerical algorithms and implementation techniques used to advance the solution. In Section~\ref{sec:tests}, we present a series of verification tests to demonstrate the code’s accuracy and robustness. In Section~\ref{sec:solarwind}, we present a global simulation of the solar wind using a simple dipolar magnetic flux distribution and a simple empirical coronal heating function \citep{Withbroe:1988} to heat the corona and drive the solar wind. The simulation produced a
solar wind and a solar corona as seen in white-light coronagraph observation qualitatively representative of the solar minimum configuration. Furthermore, the simulation suggests striking dynamic features produced by the continuous 3D reconnection in the heliospheric current sheet that can be observed by (future) white-light coronagraph observations from the poles. Finally, Section~\ref{sec:summary} gives a brief summary of the work.

\section{Governing Equations} \label{sec:equations}
We solve the semi-relativistic MHD equations in a Cartesian or spherical domain:

\begin{equation}
\frac{\partial \rho}{\partial t}=-\nabla \cdot(\rho \mathbf{v}), \label{mass_eq}
\end{equation}
\begin{equation}
\frac{\partial(\rho \mathbf{v})}{\partial t}=-\nabla \cdot(\rho \mathbf{v v})-\nabla p+\rho \mathbf{a}_g+\frac{1}{4 \pi}(\nabla \times \mathbf{B}) \times \mathbf{B} +\mathbf{F}_{\mathrm{SR}},
\label{momenta_eq}
\end{equation}
\begin{equation}
\frac{\partial \mathbf{B}}{\partial t}=-\nabla \times \mathbf{E}, \label{induction_eq}
\end{equation}
\begin{equation}
\nabla \cdot \mathbf{B}=0,
\label{zero div B}
\end{equation}
\begin{equation}
\frac{\partial e}{\partial t}=-\nabla \cdot(\mathbf{v} e)-p \nabla \cdot \mathbf{v}+\mathbf{F}_{\mathrm{NA}},
\label{energy_eq}
\end{equation}
\begin{equation}
p=\frac{\rho R T}{\mu},
\label{eq:state}
\end{equation}

where
\begin{equation}
\mathbf{E} = -\mathbf{v} \times \mathbf{B},
\end{equation}
\begin{equation}
e =\frac{p}{\gamma-1},
\end{equation}
\begin{equation}
v_A =\frac{B}{\sqrt{4 \pi \rho}},
\end{equation}
\begin{equation}
    \mathbf{F}_{\mathrm{SR}}=\frac{v_A{ }^2 / c^2}{1+v_A{ }^2 / c^2}[\mathcal{I}-\hat{\mathbf{b}} \hat{\mathbf{b}}] \cdot\left[(\rho \mathbf{v} \cdot \nabla) \mathbf{v}+\nabla p-\rho \mathbf{a}_g-\frac{1}{4 \pi}(\nabla \times \mathbf{B}) \times \mathbf{B}\right],
\end{equation}
\begin{equation}
    \mathbf{F}_{\mathrm{NA}}=-\nabla \cdot \mathbf{q}_e+Q_{\mathrm{rad}}+H_{\mathrm{em}}+H_{\mathrm{num}}.
\label{non-adibatic-terms}
\end{equation}
In the equations above, all symbols follow their conventional definitions. Specifically, $\mathbf{v}$ denotes the velocity field, $\mathbf{B}$ the magnetic field, $\mathbf{E}$ the electric field, while $\rho$, $p$, $T$ and $e$ represent plasma density, pressure,  temperature and internal energy density respectively. The term $\mathbf{F}_{\mathrm{SR}}$ appearing at the end of the momentum equation (\ref{momenta_eq}) denotes the semi-relativistic (Boris) correction term, as discussed  in \citet{boris1970physically,gombosi2002semirelativistic,rempel2016extension}, where $c$ denotes the (reduced) speed of light. $\mathbf{a}_g$ is the gravitational acceleration, $R, \mu$, and $\gamma$, are respectively the gas constant, the mean molecular weight, and the adiabatic index of the perfect gas. We assume the adiabatic index $\gamma=5 / 3$. The term $\mathbf{F}_{\mathrm{NA}}$ in the internal energy equation (\ref{energy_eq}), defined explicitly by equation (\ref{non-adibatic-terms}), accounts for the non-adiabatic processes. These include electron heat conduction, optically thin radiative cooling, empirical coronal heating, and effective heating arising from numerical diffusion in velocity and magnetic fields (from the momentum and induction equations). These non-adiabatic terms will be described in section \ref{sec:non-adiabatic}. 

\section{Numerical Algorithm} \label{sec:algorithm}

\subsection{Notations} \label{sec:notations}

In this section, we define some notations that will be frequently used throughout the paper. The MFE code employs a staggered spatial discretization as in the ZEUS code described in \citet{stone1992zeus}. In this discretization, vector quantities, including $\mathbf{v}$ (velocity) and $\mathbf{B}$ (magnetic field), are defined on the faces of finite-volume grid cells. Scalar quantities, including $\rho$ (density), $e$ (energy), $p$ (pressure), and $T$ (temperature), are defined at the centers of the finite-volume cells. Additionally, the electric field ($\mathbf{E}= -\mathbf{v} \times \mathbf{B}$) and the current density ($\nabla \times \mathbf{B}$), appearing in the induction equation, are evaluated at the edges of each volume cell.

For the coordinate system used by this code, which can be either spherical polar or Cartesian, we use $x_m$ with $m = 1, 2, 3$ to denote the coordinates corresponding to $r, \theta, \phi$ in spherical coordinates, or $x, y, z$ in Cartesian coordinates. We also utilize the following coordinate scaling coefficients (notations as in \citet{stone1992zeus}): for the spherical polar coordinate system, $g_2 = r$, $g_{31} = r$, and $g_{32} = \sin \theta$; for the Cartesian coordinate system, $g_2 = 1$, $g_{31} = 1$, and $g_{32} = 1$.

\begin{figure}[ht!]
\plotone{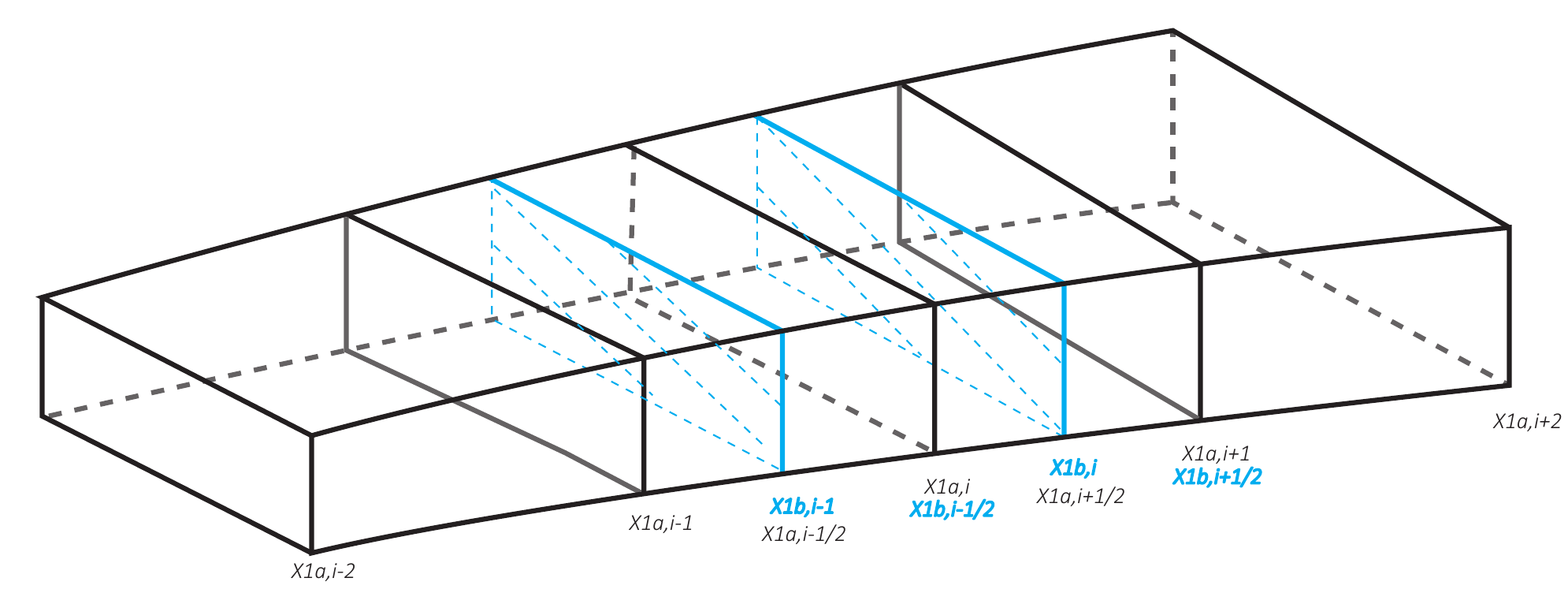}
\caption{A schematic representation of staggered grid zones in $1$-direction. The ``a-grid" locations $x_{1a,i}$ correspond to the lower interfaces $x_{1b,i-1/2}$ of the ``b-grid" zones centered on $x_{1b,i}$, 
and the ``b-grid" locations $x_{1b,i}$ correspond to the upper interfaces 
$x_{1a,i+1/2}$ of the ``a-grid" zones centered on $x_{1a,i}$
\label{fig:AB_grid}}
\end{figure}

\begin{figure}[htb!]
\centering
\includegraphics[width=0.75\textwidth]{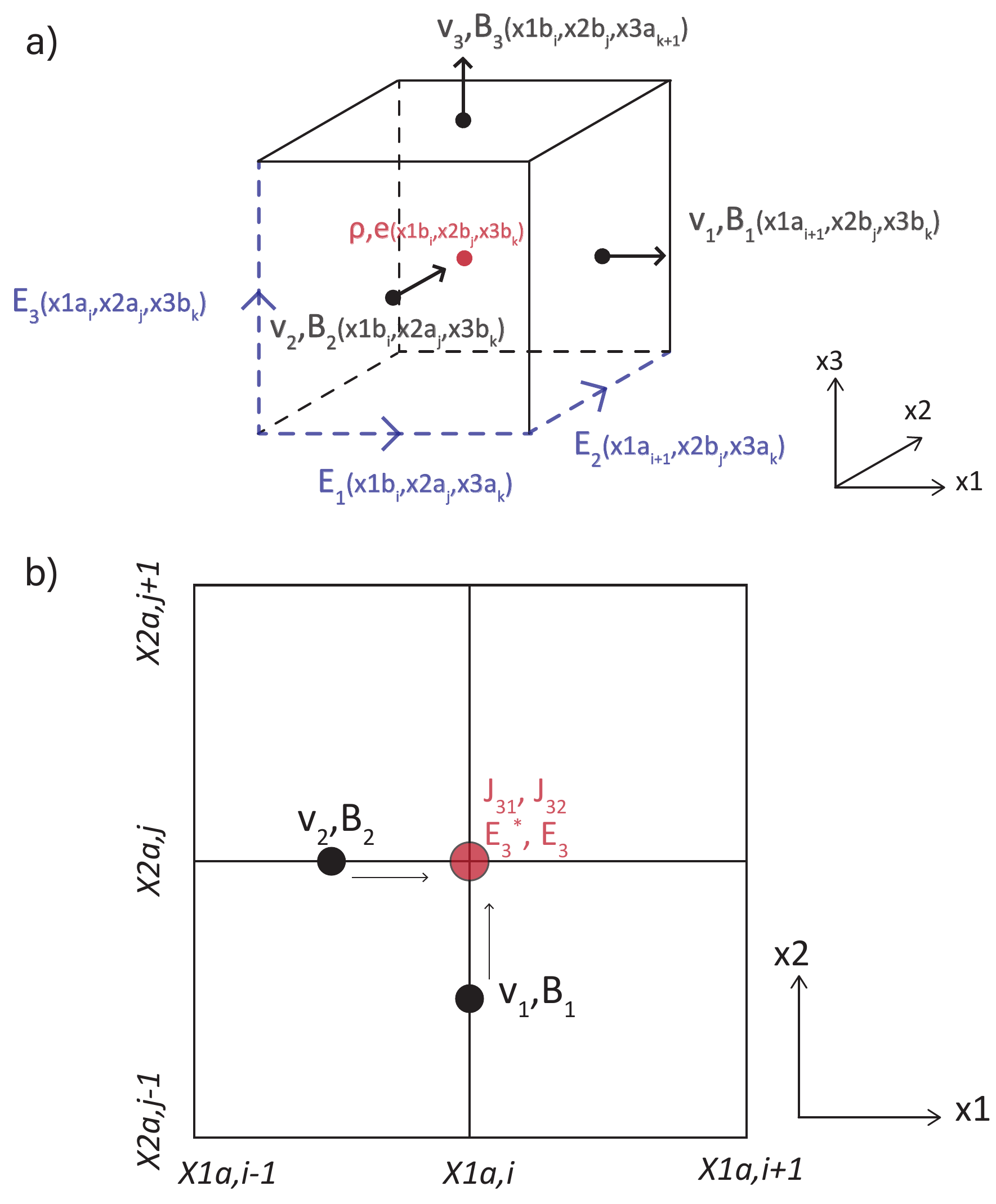}
\caption{(a) The locations of the volume centered $\rho$ and $e$, the face-centered velocity and magnetic fields, and the edge-centered electric fields in the volume cell. (b) A schematic showing in a 2D slice ($x_1$-$x_2$ plane) the evaluation of the electric fields $E_3$, $E^*_3$ and the limited current density components $J_{31}$, $J_{32}$ at the edge $\left(x_{1a,i},x_{2a,j}\right)$ using the $v_1$, $B_1$ and $v_2$, $B_2$ states reconstructed from their defined locations at the cell faces.
\label{fig:E-fields}}
\end{figure}

Subscripts $i$, $j$, and $k$ are used to represent the grid indices in the $m = 1, 2, 3$ directions, respectively.
As was done in the ZEUS code \citep{stone1992zeus}, we employ a staggered grid with two distinct grid types (``a-grid" and ``b-grid") in each spatial direction.
In 1-direction for example, the ``a-grid" with positions $x_{1a,i}$ are 
where the 1-components of the vectors ${\bf B}$ and ${\bf v}$ are defined, 
and the ``b-grid" with positions $x_{1b,i}$ centered between $x_{1a,i}$ 
and $x_{1a,i+1}$ are where the scalar quantities are defined.
Figure \ref{fig:AB_grid} shows schematically the staggard a- and b-grid zones in the 1-direction.
The a-grid locations $x_{1a,i}$ correspond to the lower interfaces 
$x_{1b,i-1/2}$ of the b-grid zones centered on $x_{1b,i}$, 
and the b-grid locations $x_{1b,i}$ correspond to the upper interfaces 
$x_{1a,i+1/2}$ of the a-grid zones centered on $x_{1a,i}$.
For a uniform grid spacing, the a-grid positions $x_{1a,i}$ are also centered between $x_{1b,i-1}$ and $x_{1b,i}$.
In the rest of the paper, coordinate position with integer index (e.g. $x_{1,i}$) refers to either the a- or b-grid zone centers (either $x_{1a,i}$ or $x_{1b,i}$) in that direction, depending upon where the considered quantity is defined; half index (e.g. $x_{1,i-1/2}$, $x_{1,i+1/2}$), accordingly represent the appropriate zone's lower and upper interfaces.
The above described notations for the grid positions in the 1-direction also apply accordingly to the 2, 3-directions with j, k indexes. In three-dimensional representation (Figure \ref{fig:E-fields}a), the scalar quantities (such as density, pressure) are centered at grid positions $\left(x_{1b,i}, x_{2b,j}, x_{3b,k}\right)$, which are the centers of the volume cells. The 1-, 2-, and 3-components of the vector ${\bf B}$ and ${\bf v}$ are centered at the volume cell faces with grid positions: $\left(x_{1a,i}, x_{2b,j}, x_{3b,k}\right)$, $\left(x_{1b,i}, x_{2a,j}, x_{3b,k}\right)$, and $\left(x_{1b,i}, x_{2b,j}, x_{3a,k}\right)$. The 1-, 2-, and 3-components of the electric field (and current density) are centered at the volume cell edges with grid positions: $\left(x_{1b,i}, x_{2a,j}, x_{3a,k}\right)$, $\left(x_{1a,i}, x_{2b,j}, x_{3a,k}\right)$, and $\left(x_{1a,i}, x_{2a,j}, x_{3b,k}\right)$.

\subsection{Reconstruction}
\label{Sec:Reconstruction}

Evaluating numerical fluxes across cell faces requires the left and right states of variables at the interfaces. In the MFE code, the reconstruction procedure interpolates cell-averaged quantities to interfaces in a manner that preserves monotonicity, thus preventing unphysical oscillations.

The reconstruction is performed in a dimensionally split manner; hence, we describe the procedure using examples in the $x_1$-dimension. To compute the states at the zone face located at $x_{1,i-\frac{1}{2}}$, which lies between the zones centered at $x_{1,i-1}$ and $x_{1,i}$, we denote the left and right values of a quantity $Q$ at the interface $x_{1,i-\frac{1}{2}}$ by $Q^L$ and $Q^R$, respectively. These interface values are obtained by reconstructing the profile of $Q$ from the zones to the left and right of the interface. To obtain at least second-order spatial accuracy, second- (or higher) order reconstruction schemes are required. We implemented the piecewise linear method (PLM) and the partial donor cell (fourth-order) method (PDM) as described in sections \ref{sec:plm} and \ref{sec:pdm} below for computing $Q^L$ and $Q^R$, because of their monotonicity preserving property and robustness in practical applications. Besides providing the left and right states, the reconstruction module supplies the following additional quantities at the interface $x_{1,i-\frac{1}{2}}$, which are subsequently used to compute advection fluxes (see Section \ref{sec:advection}):
\begin{equation} \Delta_1 Q \equiv Q^R - Q^L, \label{delta Q} \end{equation}
which represents the {\it limited} difference across the interface, and
\begin{equation} \langle Q \rangle_1 \equiv \frac{Q^R + Q^L}{2}, \label{average Q} \end{equation}
which denotes the average value at the interface. We also define the simple finite difference between the two adjacent zones:
\begin{equation}
\delta_{1} Q \equiv Q_{i} - Q_{i-1},
\label{simple_diff}
\end{equation}
as the difference between the discretized values $Q_{i-1}$, and $Q_i$ defined at  $x_{1,i-1}$ and $x_{1,i}$, which represent the zone averaged $Q$ values of the two adjacent zones.

Accordingly, $\Delta_2 Q$, $\langle Q \rangle_2$, $\delta_2 Q$ and $\Delta_3 Q$, $\langle Q \rangle_3$, $\delta_3 Q$ denote the corresponding limited difference, average value, simple finite difference obtained at the interfaces $x_{2,j-\frac{1}{2}}$ and $x_{3,k-\frac{1}{2}}$ through reconstruction along the $x_2$- and $x_3$- dimension zones, respectively.

\subsubsection{Piecewise Linear Method \label{sec:plm}}

The piecewise linear method approximates the $Q$ profile in zone $x_{1,i}$ by a linear function with a minmod slope limiter:
\begin{equation} Q(x_1) = Q_i + s_{1,i} \left( x_1 - x_{1,i} \right), \end{equation}
where the limited slope in the 1-direction, $s_{1,i}$, is given by
\begin{equation} s_{1,i} = \operatorname{minmod}\left( \frac{Q_{i+1} - Q_i}{x_{1,i+1} - x_{1,i}}, \frac{Q_i - Q_{i-1}}{x_{1,i} - x_{1,i-1}} \right). \end{equation}
The minmod function is defined as
\begin{equation}
    \operatorname{minmod} (a, b) \equiv \operatorname{sign}(a) \max [0, \min (|a|, \operatorname{sign}(a) \cdot b)].
\end{equation}
Accordingly, the left and right interface values at $x_{1,i-\frac{1}{2}}$ are computed as
\begin{equation} Q^L = Q_{i-1} + s_{1, i-1} \left( x_{1, i-1/2} - x_{1, i-1} \right), \end{equation}
\begin{equation} Q^R = Q_i - s_{1,i} \left( x_{1,i} - x_{1, i-1/2} \right). \end{equation}

\subsubsection{Partial Donor Cell (Fourth-order) Method \label{sec:pdm}}

The Partial Donor Cell (PDM) method \citep{hain1987partial} is a universal-type limiter that can be coupled with arbitrarily high-order interpolated values at the interface. Its primary idea is to constrain the solution only when the reconstructed profile may generate an imbalance in the flux of a conserved variable. When such an imbalance is detected, the limiter enforces a safe value; otherwise, the high-order interpolation is used without modification. For the left state, the general form of the PDM limiter is given by:
\begin{equation}
\begin{aligned}
Q_{}^L = {} & Q_{i-\frac{1}{2}}^{\mathrm{4th}*} - \operatorname{sign}\left(Q_{i} - Q_{i-1}\right) \cdot \max\left[ 0, \vphantom{\Big|} \right. \\
& \left. \Big|Q_{i-\frac{1}{2}}^{\mathrm{4th}*} - Q_{i-1}\Big| - A \cdot \Big|Q_{i-1} - Q_{i-2}\Big| \cdot \Big|\operatorname{sign}\left(Q_{i-1} - Q_{i-2}\right) + \operatorname{sign}\left(Q_{i} - Q_{i-1}\right)\Big| \right],
\end{aligned}
\label{PDM_general}
\end{equation}
where $Q_{i-1/2}^{\mathrm{4th}*}$ is the fourth-order interpolated value at the interface,  constrained to lie between the two adjacent cell values:
\begin{equation} Q_{i-1/2}^{\mathrm{4th}*} = \operatorname{median}\left( Q_i, Q_{i-1/2}^{\mathrm{4th}}, Q_{i-1} \right). \end{equation}
For evaluating $Q_{i-1/2}^{\mathrm{4th}}$, the coefficients for reconstruction can be derived by differentiating the Lagrange polynomial that represents the volumetric integral of the variable quantity \citep[see e.g.][]{Shu1998}. For a uniform grid, the fourth-order reconstruction weight is given as:
\begin{equation}
    Q_{i-1/2}^{\mathrm{4th}} =-\frac{1}{12}\left( Q_{i-2}+Q_{i+1}\right)+\frac{7}{12}\left( Q_{i-1}+Q_{i}\right) .
\end{equation}

The right state is obtained by reversing the reconstruction stencil:

\begin{equation}
\begin{aligned}
Q_{}^R = {} & Q_{i-\frac{1}{2}}^{\mathrm{4th}*} - \operatorname{sign}\left(Q_{i-1} - Q_{i}\right) \cdot \max\left[ 0, \vphantom{\Big|} \right. \\
& \left. \Big| Q_{i-\frac{1}{2}}^{\mathrm{4th}*}-Q_{i} \Big| - A \cdot \Big| Q_{i} -Q_{i+1} \Big| \cdot \Big|\operatorname{sign}\left(Q_{i} - Q_{i+1}\right) + \operatorname{sign}\left(Q_{i-1} - Q_{i}\right)\Big| \right] .
\end{aligned}
\label{PDM_general_R}
\end{equation}
The parameter $A$ controls the numerical diffusion: setting $A = 0$ recovers the diffusive full donor cell scheme ($Q_{}^L = Q_{i-1}$, $Q_{}^R = Q_{i}$), whereas $A > 0$ yields the partial donor cell method, effectively reducing numerical diffusion by allowing high-order interpolated values and only applying limiting based on the maximum allowed change in the donor cell controlled by $A$. In the MFE code, we set $A = 1$ according to the relation $f_{\mathrm{CFL}}<1/(1+2A)$ \citep{hain1987partial}, where $f_{\mathrm{CFL}}$ is the Courant–Friedrichs–Lewy number which is set to 0.25 in the MFE code (Section \ref{sec:time_stepping}). 

\subsection{The Update Step}\label{sec:advection}
Several important design considerations guide the numerical procedures employed in the MFE code, specifically when updating physical variables: (1) staggered grid locations require careful evaluation and reuse of numerical fluxes, (2) geometric source terms resulting from the divergence of rank-two tensors must be minimized, and (3) Lorentz forces should be formulated to minimize their field-aligned components while strictly preserving the constraint $\nabla \cdot B =0$. The explicit numerical implementations addressing these requirements are presented in coordinate form in the following sections.

\subsubsection{Continuity Equation} \label{sec:continuity_equation}
The continuity equation (\ref{mass_eq}) written explicitly in the coordinates is:
\begin{align}
    \frac{\partial \rho}{\partial t}=-\frac{1}{g_2 g_{31}} \frac{\partial}{\partial x_1}\left[g_2 g_{31}\left(\rho v_1\right)^*\right]-\frac{1}{g_2 g_{32}} \frac{\partial}{\partial x_2}\left[g_{32}\left(\rho v_2\right)^*\right]-\frac{1}{g_{31} g_{32}} \frac{\partial}{\partial x_3}\left(\rho v_3\right)^*  ,  \label{discretized rho}
\end{align}
where the fluxes with superscript `*' denote an upwinded evaluation of the fluxes at the appropriate cell interfaces. We use a modified Lax-Friedrichs scheme \citep{rempel2009radiative} to obtain the upwinded evaluation of these fluxes as follows. As an example, consider the upwinded evaluation of the 1-component of the mass flux through the density cell interface at $x_{1b,i-1/2}$:
\begin{equation}
    \left(\rho v_1\right)_{i-1 / 2}^*=\left(v_1\right)_{i-1 / 2} \langle \rho \rangle_1-\left(\frac{\left|v_1\right|+c_f q_{1, \rho}^l}{2}\right)_{i-1 / 2} \Delta_1 \rho.
    \label{mass flux1}
\end{equation}
At the cell interface for density the velocity $\left(v_1\right)_{i-1 / 2}$ is already defined there, while the density left and right states $\rho^L$ and $\rho^R$ at the interface $x_{1b, i-1/2}$ require reconstruction. The term $\langle \rho \rangle_1$ as given by equation (\ref{average Q}) corresponds to the average of the density left right states at the interface through reconstruction along the $x_1$-dimension, and $\Delta_1 \rho$ as given by equation (\ref{delta Q}) corresponds to the limited difference of the density left right states at the interface. The factor $q_{1, \rho}^l$ is used to limit the upwinding fast mode speed $c_f$, and is given by:
\begin{equation}
    q_{1, \rho}=\frac{\Delta_1 \rho}{\delta_1 \rho}
    \label{q_den}
\end{equation}
with $\delta_1 \rho$ being the simple finite difference of density between the two adjacent zones in the $x_1$-dimension as given by equation (\ref{simple_diff}). We note that because of the monotonicity preserving reconstruction used, it is guaranteed that this ratio of the limited difference over the finite difference satisfies $0 \le q_{1, \rho} \leq 1$, and becomes significantly less than 1 when the profile is smooth. Thus, $q_{1, \rho}^l$, which is $q_{1,\rho}$ raised to the power of $l$ becomes $\ll 1$ when the profile is smooth, reducing the contribution from the fast mode speed $c_f$ to the diffusive flux in the last term in equation (\ref{mass flux1}). Conversely, when discontinuities occur, the reconstruction slopes flatten, causing the limited difference to approach the simple finite difference and thus $q_{1, \rho}=\left(\Delta_1 \rho\right)/\left(\delta_1 \rho\right)=1$. In this limit, the diffusive flux uses the full speed, as in the standard Lax-Friedrichs method. In practice, we find that using $l=4$ with PLM is a good choice for reducing numerical diffusion while maintaining robustness. We use $l=0$ with PDM because the PDM reconstruction already gives a very small left-right state difference (hence small diffusion) because of the high-order reconstruction, that using higher $l$ can be over-aggressive.  Therefore we adopt the standard Lax-Friedrichs scheme with $q_{1, \rho}^l=1$ always holds for the PDM reconstruction.

Similarly, the 2- and 3- components of the mass flux are calculated as follows:
\begin{equation}
    \left(\rho v_2\right)_{j-1 / 2}^*=\left(v_2\right)_{j-1 / 2} \langle \rho \rangle_2-\left(\frac{\left|v_2\right|+c_f q_{2, \rho}^l}{2}\right)_{j-1 / 2} \Delta_2 \rho,
    \label{mass flux2}
\end{equation}
and
\begin{equation}
    \left(\rho v_3\right)_{k-1 / 2}^*=\left(v_3\right)_{k-1 / 2} \langle \rho \rangle_3-\left(\frac{\left|v_3\right|+c_f q_{3, \rho}^l}{2}\right)_{k-1 / 2} \Delta_3 \rho,
    \label{mass flux3}
\end{equation}
at the density cell interfaces at $x_{2b,j-1/2}$ and $x_{3b, k-1/2}$ respectively.  

\subsubsection{Energy Equation}

The internal energy equation written explicitly in the coordinates is
\begin{align}
\frac{\partial e}{\partial t}= & -\frac{1}{g_2 g_{31}} \frac{\partial}{\partial x_1}\left[g_2 g_{31}\left(\rho v_1 \frac{e}{\rho}\right)^*\right]-\frac{1}{g_2 g_{32}} \frac{\partial}{\partial x_2}\left[g_{32}\left(\rho v_2 \frac{e}{\rho}\right)^*\right]-\frac{1}{g_{31} g_{32}} \frac{\partial}{\partial x_3}\left[ \left(\rho v_3 \frac{e}{\rho}\right)^* \right] \notag\\
& -p\left[\frac{1}{g_2 g_{31}} \frac{\partial}{\partial x_1}\left(g_2 g_{31} v_1\right)+\frac{1}{g_2 g_{32}} \frac{\partial}{\partial x_2}\left(g_{32} v_2\right)+\frac{1}{g_{31} g_{32}} \frac{\partial v_3}{\partial x_3}\right]  + \mathbf{F}_{\mathrm{na}}, \label{discretized energy}
\end{align}
where $\mathbf{F}_{\mathrm{na}}$ represents the non-adiabatic terms. The fourth term on the right hand side (RHS) of equation (\ref{discretized energy}) represents $p \nabla \cdot \mathbf{v}$, evaluated at the volume cell center location of $e$ by carrying out simple centered (second-order) finite differences of $v_1$, $v_2$ and $v_3$ defined on the respective cell faces. No interpolation adjustments are needed. The starred (`$^*$') fluxes in the advection terms are computed through reconstruction and upwinding as described in the following. Instead of directly reconstructing the energy $e$, we choose to reconstruct the specific internal energy $e/\rho$. The upwinded 1-flux $\rho v_1(e / \rho)$ through the $e$ (or $\rho$) cell interface at $x_{1b,i-1/2}$ is:
\begin{equation}
    \left(\rho v_1 \frac{e}{\rho}\right)_{i-1 / 2}^*=  \left(\rho v_1\right)_{i-1 / 2} \langle \frac{e}{\rho} \rangle_1  -\left(\frac{\left|\rho v_1\right|+\rho c_f q_{1, e}^l}{2}\right)_{i-1 / 2} \Delta_1\left(\frac{e}{\rho}\right)
\end{equation}
where we use the mass flux evaluated in equation (\ref{mass flux1}) for the term $\left(\rho v_1\right)_{i-1 / 2}$ here. The term $ \langle e/\rho \rangle_1$ as given by equation (\ref{average Q}) corresponds to the average of the left right states of $e/\rho$ at the interface $x_{1b,i-1/2}$, and $\Delta_1 \left(e/\rho\right)$ as given by equation (\ref{delta Q}) corresponds to the limited difference of the left right $e/\rho$ states, and $q_{1, e}=\frac{\Delta_1 \left( e/\rho\right)}{\delta_1 \left( e/\rho \right) }$, which is the ratio between the limited difference and the simple finite difference (eq. [\ref{simple_diff}]) of $e/\rho$, is the factor used to limit the effective upwind speed.

Similarly, the 2- and 3- components of the upwinded energy flux are calculated as follows:
\begin{equation}
    \left(\rho v_2 \frac{e}{\rho}\right)_{j-1 / 2}^*=  \left(\rho v_2\right)_{j-1 / 2} \langle \frac{e}{\rho} \rangle_2 
 -\left(\frac{\left|\rho v_2\right|+\rho c_f q_{2, e}^l}{2}\right)_{j-1 / 2} \Delta_2\left(\frac{e}{\rho}\right)
\end{equation}
and
\begin{equation}
    \left(\rho v_3\frac{e}{\rho}\right)_{k-1 / 2}^*= \left(\rho v_3\right)_{k-1 / 2} \langle \frac{e}{\rho} \rangle_3  -\left(\frac{\left|\rho v_3\right|+\rho c_f q_{3, e}^l}{2}\right)_{k-1 / 2} \Delta_3\left(\frac{e}{\rho}\right)
\end{equation}
at the $e$ (or $\rho$) cell interfaces at $x_{2b,j-1/2}$, $x_{3b,k-1/2}$ respectively.

\subsubsection{Momentum Equation}
\label{Sec:momenta}
Making use of the continuity equation and the fact that the Lorentz force
term is always perpendicular to ${\bf B}$, the momentum equation
(\ref{momenta_eq}) can be rewritten as:
\begin{eqnarray}
\frac{\partial ( {\rho \bf v})}{\partial t}
&=& - \nabla \cdot \left ( \rho {\bf v} {\bf v} \right )
- \nabla p + \rho {\bf a}_g
\nonumber \\
& & - \frac{{v_A}^2/c^2}{1 + {v_A}^2/c^2} \left[ {\cal I}
- {\hat{\bf b}} {\hat{\bf b}} \right ]
\cdot \left [ - \nabla \cdot \left ( \rho {\bf v} {\bf v} \right )
- \nabla p + \rho {\bf a}_g -{\bf v} \frac{\partial \rho}{\partial t}
\right ]
\nonumber \\
& & + \frac{1}{1 + {v_A}^2/c^2}\frac{1}{4 \pi} ( \nabla \times {\bf B} ) \times {\bf B},
\nonumber \\
&=& {\bf F}_h  - \frac{{v_A}^2/c^2}{1 + {v_A}^2/c^2} \left[ {\cal I}
- {\hat{\bf b}} {\hat{\bf b}} \right ] \cdot \left [{\bf F}_h - {\bf v} \frac{\partial \rho}{\partial t}
\right ] + \frac{1}{1 + {v_A}^2/c^2} {\bf F}_L,
\label{eq:motion1}
\end{eqnarray}
in which ${\bf F}_h \equiv - \nabla \cdot \left ( \rho {\bf v} {\bf v} \right )
- \nabla p + \rho {\bf a}_g$ and
${\bf F}_L \equiv (1/4 \pi) \, ( \nabla \times {\bf B} ) \times {\bf B}$
denote the classic MHD hydrodynamic and the Lorentz force terms.
Writing the equation explicitly in the coordinates:
\begin{equation}
\frac{\partial}{\partial t}\left(\rho v_1\right) =
F_{h 1} - \frac{{v_A}^2/c^2}{1 + {v_A}^2/c^2} \left[F_{h 1} - b_1 ({\hat{\bf b}} \cdot {\bf F}_h) - (v_1 - b_1 ({\hat{\bf b}} \cdot {\bf v})) \frac{\partial \rho}{\partial t} \right] + \frac{1}{1 + {v_A}^2/c^2} F_{L 1},
\label{eq:mom1}
\end{equation}
\begin{equation}
\frac{\partial}{\partial t}\left(\rho v_2\right) =
F_{h 2} - \frac{{v_A}^2/c^2}{1 + {v_A}^2/c^2} \left[F_{h 2} - b_2 ({\hat{\bf b}} \cdot {\bf F}_h) - (v_2 - b_2 ({\hat{\bf b}} \cdot {\bf v})) \frac{\partial \rho}{\partial t} \right] + \frac{1}{1 + {v_A}^2/c^2} F_{L 2},
\label{eq:mom2}
\end{equation}
\begin{equation}
\frac{\partial}{\partial t}\left(\rho v_3\right) =
F_{h 3} - \frac{{v_A}^2/c^2}{1 + {v_A}^2/c^2} \left[F_{h 3} - b_3 ({\hat{\bf b}} \cdot {\bf F}_h) - (v_3 - b_3 ({\hat{\bf b}} \cdot {\bf v})) \frac{\partial \rho}{\partial t} \right] + \frac{1}{1 + {v_A}^2/c^2} F_{L 3},
\label{eq:mom3}
\end{equation}
where ${\hat{\bf b}}={\bf B}/B$ is the unit vector of ${\bf B}$ with $b_1, b_2, b_3$ denoting its 3 components, and
\begin{align}
    F_{h 1} &=   -\frac{1}{g_2 g_{31}} \frac{\partial}{\partial x_1}\left[g_2 g_{31}\left(\rho v_1 v_1\right)^*\right] 
 -\frac{1}{g_2 g_{32}} \frac{\partial}{\partial x_2}\left[g_{32}\left(\rho v_2 v_1\right)^*\right] 
 -\frac{1}{g_{31} g_{32}} \frac{\partial}{\partial x_3}\left(\rho v_3 v_1\right)^*  \notag \\
  & \quad \quad + \rho v_2^2 \frac{1}{g_2} \frac{\partial g_2}{\partial x_1}+\rho v_3^2 \frac{1}{g_{31}} \frac{\partial g_{31}}{\partial x_1} 
 -\frac{\partial p}{\partial x_1} + \rho a_{g1} , \label{F_h1} \\
 F_{h 2} &=  -\frac{1}{g_2^2 g_{31}} \frac{\partial}{\partial x_1}\left[g_2^3 g_{31} \left(\rho v_1 \frac{v_2}{g_2}\right)^*\right] 
-  \frac{1}{g_2^2 g_{32}} \frac{\partial}{\partial x_2}\left[g_2^2 g_{32} \left(\rho v_2 \frac{v_2}{g_2}\right)^*\right] \notag \\
 & \quad \quad -\frac{1}{g_2 g_{31} g_{32}} \frac{\partial}{\partial x_3}\left[g_2^2\left(\rho v_3 \frac{v_2}{g_2}\right)^*\right] \notag\\
 & \quad \quad +  \rho v_3^2 \frac{1}{g_{32} g_2} \frac{\partial g_{32}}{\partial x_2}  -\frac{1}{g_2} \frac{\partial p}{\partial x_2} + \rho a_{g2}, \label{F_h2} \\
 F_{h 3} & =-\frac{1}{g_2 g_{31}^2 g_{32}} \frac{\partial}{\partial x_1}\left[g_2 g_{31}^3 g_{32}^2\left(\rho v_1 \frac{v_3}{g_{31} g_{32}}\right)^*\right] \notag \\
 & \quad \quad -\frac{1}{g_2 g_{31} g_{32}^2 } \frac{\partial}{\partial x_2}\left[g_{31}^2 g_{32}^3\left(\rho v_2 \frac{v_3}{g_{31} g_{32}}\right)^*\right] \notag \\
& \quad \quad -\frac{1}{g_{31}^2 g_{32}^2 } \frac{\partial}{\partial x_3}\left[g_{31}^2 g_{32}^2\left(\rho v_3 \frac{v_3}{g_{31} g_{32}}\right)^*\right] \notag \\
& \quad \quad -\frac{1}{g_{31} g_{32}} \frac{\partial p}{\partial x_3} + \rho a_{g3}, \label{F_h3}
\end{align}
\begin{align}
 F_{L 1} &= \frac{1}{4 \pi} (j_2 B_3 - j_3 B_2) \nonumber \\
 &= \frac{1}{4 \pi} \left[ \frac{B_3}{g_{31} g_{32}} \frac{\partial B_1}{\partial x_3}-\frac{B_3}{g_{31}} \frac{\partial}{\partial x_1}\left(g_{31} B_3\right)-\frac{B_2}{g_2} \frac{\partial}{\partial x_1}\left(g_2 B_2\right)+\frac{B_2}{g_2} \frac{\partial B_1}{\partial x_2} \right], \label{F_L1}\\
 F_{L 2} &= \frac{1}{4 \pi} (j_3 B_1 - j_1 B_3) \nonumber \\
 & = \frac{1}{4 \pi} \left[ \frac{B_1}{g_2} \frac{\partial}{\partial x_1}\left(g_2 B_2\right)-\frac{B_1}{g_2} \frac{\partial B_1}{\partial x_2}-\frac{B_3}{g_{32} g_2} \frac{\partial}{\partial x_2}\left(g_{32} B_3\right)+\frac{B_3}{g_{31} g_{32}} \frac{\partial B_2}{\partial x_3} \right], \label{F_L2}\\
 F_{L 3} &= \frac{1}{4 \pi} (j_1 B_2 - j_2 B_1) \nonumber \\
 & = \frac{1}{4 \pi} \left[ \frac{B_2}{g_{32} g_2} \frac{\partial}{\partial x_2}\left(g_{32} B_3\right)-\frac{B_2}{g_{31} g_{32}} \frac{\partial B_2}{\partial x_3}-\frac{B_1}{g_{31} g_{32}} \frac{\partial B_1}{\partial x_3}+\frac{B_1}{g_{31}} \frac{\partial}{\partial x_1}\left(g_{31} B_3\right) \right]. \label{F_L3}
\end{align}
In equations (\ref{F_h1})-(\ref{F_h3}), $a_{g1}$, $a_{g2}$, and $a_{g3}$ are the three components of the gravitational acceleration, and in the case of spherical geometry, only $a_{g1}$ (in the $r$-direction) is non-zero. The Lorentz force components in equations (\ref{F_L1})-(\ref{F_L3}) are given directly in the form of $(1/4 \pi) \, {\bf j} \times {\bf B}$ with ${\bf j} \equiv \nabla \times {\bf B}$, instead of using the conservative form of the divergence of the Maxwell stress, to minimize the error of field aligned force. Note that using the latter conservative form would effectively introduce extra terms parallel to ${\bf B}$ that are proportional to $\nabla \cdot {\bf B}$, which result in a significant error of field aligned force if there is significant non-zero $\nabla \cdot {\bf B}$ error present.
Also note in equation (\ref{F_h2}) the geometric source term $\frac{\rho v_1 v_2}{g_{2}} \frac{\partial g_{2}}{\partial x_1}$ is absorbed by the first term on the right hand side, and equation (\ref{F_h3}) is the so-called  $\phi$-angular momentum conservation form \citep{kley1998treatment}.

The term $\partial \rho / \partial t$ in the right hand sides of equations
(\ref{eq:mom1}), (\ref{eq:mom2}), and (\ref{eq:mom3}) is evaluated by the continuity equation as described in (section \ref{sec:continuity_equation}). For evaluating $F_{h 1}$, $F_{h 2}$, $F_{h 3}$ given by equations (\ref{F_h1}) to (\ref{F_h3}), direct linear interpolations and second order finite-differences are applied to all the quantities and derivatives, except for the fluxes in $(\quad)^*$, where the upwinded evaluations are performed at the appropriate cell faces as described in the following.

For the first three terms on the RHS of equation (\ref{F_h1}), the upwinded evaluation of the 1-, 2-, and 3-fluxes $\rho v_1 v_1,\rho v_2 v_1,\rho v_3 v_1$ through the $v_1$ cell faces at respectively $x_1=x_{1a,i-1/2},x_2=x_{2b,j-1/2},x_3=x_{3b,k-1/2}$ are: 
\begin{equation}
\left(\rho v_1 v_1\right)_{i-1 / 2}^*=\left(\rho v_1\right)_{i-1 / 2} \langle v_1 \rangle_1-\left(\frac{\left|\rho v_1\right|+\rho c_f q_{1, v_1}^l}{2}\right)_{i-1 / 2} \Delta_1 v_1, \label{m1 flux1}
\end{equation}
\begin{equation}
\left(\rho v_2 v_1\right)_{j-1 / 2}^*=\left(\rho v_2\right)_{j-1 / 2} \langle v_1 \rangle_2-\left(\frac{\left|\rho v_2\right|+\rho c_f q_{2, v_1}^l}{2}\right)_{j-1 / 2} \Delta_2 v_1, \label{m1 flux2}
\end{equation}
\begin{equation}
\left(\rho v_3 v_1\right)_{k-1 / 2}^*=\left(\rho v_3\right)_{k-1 / 2} \langle v_1 \rangle_3-\left(\frac{\left|\rho v_3\right|+\rho c_f q_{3, v_1}^l}{2}\right)_{k-1 / 2} \Delta_3 v_1, \label{m1 flux3}
\end{equation}
where $\left(\rho v_1\right)_{i-1 / 2},\left(\rho v_2\right)_{j-1 / 2},\left(\rho v_3\right)_{k-1 / 2}$ are obtained by averaging the mass fluxes evaluated in equations (\ref{mass flux1}), (\ref{mass flux2}), (\ref{mass flux3}). The term $\langle v_{1} \rangle_1$, $\langle v_{1} \rangle_2$, $\langle v_{1} \rangle_3$ correspond to the averages (eq. [\ref{average Q}]) of the $v_1$ left right states evaluated at the respective cell faces at $x_1=x_{1a,i-1/2},x_2=x_{2b,j-1/2},x_3=x_{3b,k-1/2}$, and $\Delta_1 v_{1}$, $\Delta_2 v_{1}$, $\Delta_3 v_{1}$ correspond to the limited differences (eq. [\ref{delta Q}]) of the $v_1$ left right states evaluated at the respective cell faces. $q_{m, v_1}=\frac{\Delta_m v_1}{\delta_m v_1}$ with $m=1,2,3$ are the ratios of the limited difference over the simple finite difference (eq. [\ref{simple_diff}]) at the respective cell faces, which provide the limiting factors for the upwind speed.

In the same way, for equation (\ref{F_h2}), the upwinded 1-, 2-, and 3-fluxes $\rho v_1\left(v_2 / g_2\right), \rho v_2\left(v_2 / g_2\right)$, and $\rho v_3\left(v_2 / g_2\right)$ through the $v_2$ cell faces at respectively $x_1= x_{1b,i-1/2}$, $x_2=x_{2a,j-1/2}$, $x_3=x_{3b,k-1/2}$ are:
\begin{equation}
    \left(\rho v_1 \frac{v_2}{g_2}\right)_{i-1 / 2}^*=\left(\rho v_1\right)_{i-1 / 2} \langle \frac{v_2}{g_2} \rangle_1-\left(\frac{\left|\rho v_1\right|+\rho c_f q_{1, v_2}^l}{2}\right)_{i-1 / 2} \Delta_1\left(\frac{v_2}{g_2}\right), \label{m2 flux1}
\end{equation}
\begin{equation}
\left(\rho v_2 \frac{v_2}{g_2}\right)_{j-1 / 2}^*=\left(\rho v_2\right)_{j-1 / 2} \langle\frac{v_2}{g_2} \rangle_2-\left(\frac{\left|\rho v_2\right|+\rho c_f q_{2, v_2}^l}{2}\right)_{j-1 / 2} \Delta_2\left(\frac{v_2}{g_2}\right), \label{m2 flux2}
\end{equation}
\begin{equation}
\left(\rho v_3 \frac{v_2}{g_2}\right)_{k-1 / 2}^*=\left(\rho v_3\right)_{k-1 / 2} \langle\frac{v_2}{g_2} \rangle_3-\left(\frac{\left|\rho v_3\right|+\rho c_f q_{3, v_2}^l}{2}\right)_{k-1 / 2} \Delta_3\left(\frac{v_2}{g_2}\right), \label{m2 flux3}
\end{equation}
where $q_{m, v_2}=\frac{\Delta_m\left(v_2 / g_2\right)}{\delta_m\left(v_2 / g_2\right)}$, with $m=1,2,3$.

And for equation (\ref{F_h3}), the upwinded 1-, 2-, and 3-fluxes $\rho v_1\left(v_3 / g_{31} g_{32}\right), \rho v_2\left(v_3 / g_{31} g_{32}\right)$, and $\rho v_3\left(v_3 / g_{31} g_{32}\right)$ through the $v_3$ cell faces at respectively $x_1=x_{1b,i-1/2}$, $x_2=x_{2b,j-1/2}$, $x_3=x_{3a,k-1/2}$ are:
\begin{equation}
\left(\rho v_1 \frac{v_3}{g_{31} g_{32}}\right)_{i-1 / 2}^*=\left(\rho v_1\right)_{i-1 / 2} \langle \frac{v_3}{g_{31} g_{32}} \rangle_1-\left(\frac{\left|\rho v_1\right|+\rho c_f q_{1, v_3}^l}{2}\right)_{i-1 / 2} \Delta_1\left(\frac{v_3}{g_{31} g_{32}}\right), \label{m3 flux1}
\end{equation}
\begin{equation}
\left(\rho v_2 \frac{v_3}{g_{31} g_{32}}\right)_{j-1 / 2}^*=\left(\rho v_2\right)_{j-1 / 2} \langle \frac{v_3}{g_{31} g_{32}} \rangle_2-\left(\frac{\left|\rho v_2\right|+\rho c_f q_{2, v_3}^l}{2}\right)_{j-1 / 2} \Delta_2\left(\frac{v_3}{g_{31} g_{32}}\right), \label{m3 flux2}
\end{equation}
\begin{equation}
\left(\rho v_3 \frac{v_3}{g_{31} g_{32}}\right)_{k-1 / 2}^*=\left(\rho v_3\right)_{k-1 / 2} \langle \frac{v_3}{g_{31} g_{32}} \rangle_3-\left(\frac{\left|\rho v_3\right|+\rho c_f q_{3, v_3}^l}{2}\right)_{k-1 / 2} \Delta_3\left(\frac{v_3}{g_{31} g_{32}}\right), \label{m3 flux3}
\end{equation}
where $q_{m, v_3}=\frac{\Delta_m\left(v_3 / g_{31}g_{32}\right)}{\delta_m\left(v_3 / g_{31}g_{32}\right)}$, with $m=1,2,3$.

For evaluating the the Lorentz force components $F_{L1}$, $F_{L2}$, $F_{L3}$ given by equations (\ref{F_L1})-(\ref{F_L3}) in the form of ${\bf j} \times {\bf B}$, we first evaluate them centered at the volume cell edges where the necessary ${\bf j}$ components are naturally defined and evaluated by simple second order finite differences of the face-centered magnetic field components. The necessary magnetic field components are also linearly interpolated to the cell edges for computing the edge-centered ${\bf j} \times {\bf B}$ Lorentz force components. The computed edge-centered Lorentz force components are then linearly interpolated to the cell faces to obtain the final face-centered ${\bf F}_{L}$ components needed for the momentum update.

\subsubsection{Constrained Transport}
\label{Sec:E-fields}

For solving the induction equation (\ref{induction_eq}) we use the constrained transport (CT) scheme \citep{evans1988simulation} on the staggered grid to ensure the divergence free condition for the magnetic field (eq. [\ref{zero div B}]) is satisfied to round-off errors. By applying Stokes' Theorem, the CT scheme evaluates the $\mathbf{E}$ field in equation (\ref{induction_eq}) at the volume cell edges that enclose the volume cell faces where magnetic fields are located (see Figure \ref{fig:E-fields}). The electric field components cancel pairwise during the update of the magnetic flux integrated over all the faces of the volume cell, thus the $\nabla \cdot B =0$ is conserved in the integral sense.
The induction equation we solve written explicitly in the coordinates are:
\begin{equation}
    \frac{\partial B_1}{\partial t} = -\frac{1}{g_2 g_{31} g_{32} } \left[ \frac{\partial}{\partial x_2}\left(g_{31} g_{32} E_3 \right)
    - \frac{\partial}{\partial x_3}\left(g_{2} E_2 \right) \right],
\label{eq:B1}
\end{equation}
\begin{equation}
    \frac{\partial B_2}{\partial t} = -\frac{1}{ g_{31} g_{32} } \left[ \frac{\partial}{\partial x_3}\left( E_1 \right)
    - \frac{\partial}{\partial x_1}\left(g_{31} g_{32} E_3 \right) \right],
\label{eq:B2}
\end{equation}
\begin{equation}
    \frac{\partial B_3}{\partial t} = -\frac{1}{g_2 } \left[ \frac{\partial}{\partial x_1}\left(g_{2} E_2 \right)
    - \frac{\partial}{\partial x_2}\left( E_1 \right) \right] 
\label{eq:B3}
\end{equation}
where $B_1$, $B_2$, and $B_3$ are located at the volume cell faces with positions given by $\left(x_{1a,i},x_{2b,j},x_{3b,k}\right) $, $\left(x_{1b,i},x_{2a,j},x_{3b,k}\right)$ and $\left(x_{1b,i},x_{2b,j},x_{3a,k}\right)$ respectively (Figure \ref{fig:E-fields}). $E_1$, $E_2$, and $E_3$ are located at the volume cell edges with the positions given by  $\left(x_{1b,i},x_{2a,j},x_{3a,k}\right) $, $\left(x_{1a,i},x_{2b,j},x_{3a,k}\right)$ and $\left(x_{1a,i},x_{2a,j},x_{3b,k}\right)$ respectively (Figure \ref{fig:E-fields}), and are evaluated as follows:
\begin{equation}
    E_{1}= -\left[ \left<v_2 \right>_{3} \left<B_3\right>_{2} - \left<v_3\right>_{2} \left<B_2\right>_{3} - {E_1}^* \right]
\end{equation}
\begin{equation}
    E_{2}= -\left[ \left<v_3\right>_{1} \left<B_1\right>_{3} - \left<v_1\right>_{3} \left<B_3\right>_{1} - {E_2}^* \right]
\end{equation}
\begin{equation}
    E_{3}= -\left[ \left<v_1\right>_{2} \left<B_2\right>_{1} - \left<v_2\right>_{1} \left<B_1\right>_{2} -{E_3}^* \right].
\end{equation}
The terms in $\langle \quad \rangle_m$ with $m=1,2,3$ represent the averages (as defined in equation [\ref{average Q}]) of the reconstructed left and right states of the velocity and magnetic fields at the cell edges, which comprise the convective electric field. The ${E_1}^*$, ${E_2}^*$, ${E_3}^*$ are the diffusive electric field components that provide the appropriate upwinding for the convective electric field and are given by:
\begin{equation}
{E_1}^* = \eta_{12} J_{12} - \eta_{13} J_{13},
\label{diffemf1}
\end{equation}
\begin{equation}
{E_2}^* = \eta_{23} J_{23} - \eta_{21} J_{21},
\label{diffemf2}
\end{equation}
\begin{equation}
{E_3}^* = \eta_{31} J_{31} - \eta_{32} J_{32},
\label{diffemf3}
\end{equation}
where
\begin{equation}
    J_{12}=\frac{\Delta_2\left(B_3 g_{31} g_{32} d x_3\right)}{g_2 g_{31} g_{32} d x_2 d x_3},
\label{limited J12}
\end{equation}
\begin{equation}
    J_{13} = \frac{\Delta_3\left(B_2 g_2 d x_2\right)}{g_2 g_{31} g_{32} d x_2 d x_3},
\label{limited J13}
\end{equation}
\begin{equation}
    J_{23}=\frac{\Delta_3\left(B_1 d x_1\right)}{g_{31} g_{32} d x_1 d x_3}
\label{limited J23}
\end{equation}
\begin{equation}
    J_{21}=\frac{\Delta_1\left(B_3 g_{31} g_{32} d x_3\right)}{g_{31} g_{32} d x_1 d x_3},
\label{limited J21}
\end{equation}
\begin{equation}
    J_{31}= \frac{\Delta_1\left(B_2 g_2 d x_2\right)}{g_2 d x_1 d x_2},
\label{limited J31}
\end{equation}
\begin{equation}
    J_{32}=\frac{\Delta_2\left(B_1 d x_1\right)}{g_2 d x_1 d x_2},
\label{limited J32}
\end{equation}
which are effectively the various components of the current density computed at the appropriate cell edges using the limited differences (see the definition of $\Delta_m$ in equation [\ref{delta Q}]) of the reconstructed left right states of the ${\bf B}$ field components times the geometric factors. In the above, $dx_1$, $dx_2$, $dx_3$ denote the cell sizes in the 1, 2, 3-directions. Figure \ref{fig:E-fields}b illustrates the locations $J_{31}$,$J_{32}$ and $E_3$ are evaluated. The magnitudes of these limited current density components are generally significantly smaller than those of the actual current density computed with direct second order finite differences. The $\eta_{12}$, $\eta_{13}$, $\eta_{23}$, $\eta_{21}$, $\eta_{31}$, $\eta_{32}$ are the local numerical diffusion coefficients, which when PLM is used are given as follows:
\begin{equation}
\eta_{12} = \min \left( \frac{1}{2} \frac{(g_2 dx_2)^2}{\sqrt{\rho}} \left| J_{12} \right|, \frac{1}{2} c \, g_2 dx_2 \right) q_{12}^l + \frac{1}{2} \left| v_2\right| g_2 dx_2
\label{eta12}
\end{equation}
\begin{equation}
\eta_{13}=\min \left( \frac{1}{2} \frac{(g_{31} g_{32} dx_3)^2}{\sqrt{\rho}} \left| J_{13}  \right|, \frac{1}{2} c \, g_{31} g_{32} dx_3 \right) q_{13}^l + \frac{1}{2} \left| v_3\right| g_{31} g_{32} dx_3
\label{eta13}
\end{equation}
\begin{equation}
\eta_{23}=\min \left( \frac{1}{2} \frac{(g_{31} g_{32} dx_3)^2}{\sqrt{\rho}} \left| J_{23} \right|, \frac{1}{2} c \, g_{31} g_{32} dx_3 \right) q_{23}^l + \frac{1}{2} \left| v_3\right| g_{31} g_{32} dx_3
\label{eta23}
\end{equation}
\begin{equation}
\eta_{21}=\min \left( \frac{1}{2} \frac{(dx_1)^2}{\sqrt{\rho}} \left| J_{21}  \right|, \frac{1}{2} c \, dx_1 \right) q_{21}^l + \frac{1}{2} \left| v_1\right| dx_1
\label{eta21}
\end{equation}
\begin{equation}
\eta_{31}=\min \left( \frac{1}{2} \frac{(dx_1)^2}{\sqrt{\rho}} \left| J_{31} \right|, \frac{1}{2} c \, dx_1 \right) q_{31}^l + \frac{1}{2} \left| v_1 \right| dx_1
\label{eta31}
\end{equation}
\begin{equation}
\eta_{32}=\min \left( \frac{1}{2} \frac{(g_2 dx_2)^2}{\sqrt{\rho}} \left| J_{32}  \right|, \frac{1}{2} c \, g_2 dx_2 \right) q_{32}^l + \frac{1}{2} \left| v_2\right| g_2 dx_2,
\label{eta32}
\end{equation}
for which the first terms on the RHS are themselves proportional to the magnitude of the limited current density components, and are further limited by multiplying by the factors (raised to the $l$th power):
\begin{equation}
q_{12} = J_{12}/j_{12}
\label{q12}
\end{equation}
\begin{equation}
q_{13}=J_{13}/j_{13}
\label{q13}
\end{equation}
\begin{equation}
q_{23}=J_{23}/j_{23}
\label{q23}
\end{equation}
\begin{equation}
q_{21}=J_{21}/j_{21}
\label{q21}
\end{equation}
\begin{equation}
q_{31}=J_{31}/j_{31}
\label{q31}
\end{equation}
\begin{equation}
q_{32}=J_{32}/j_{32}
\label{q32}
\end{equation}
with
\begin{equation}
    j_{12}=\frac{\delta_2\left(B_3 g_{31} g_{32} d x_3\right)}{g_2 g_{31} g_{32} d x_2 d x_3}
\label{simple J12}
\end{equation}
\begin{equation}
    j_{13} = \frac{\delta_3\left(B_2 g_2 d x_2\right)}{g_2 g_{31} g_{32} d x_2 d x_3}
\label{simple J13}
\end{equation}
\begin{equation}
    j_{23}=\frac{\delta_3\left(B_1 d x_1\right)}{g_{31} g_{32} d x_1 d x_3}
\label{simple J23}
\end{equation}
\begin{equation}
    j_{21}=\frac{\delta_1\left(B_3 g_{31} g_{32} d x_3\right)}{g_{31} g_{32} d x_1 d x_3}
\label{simple J21}
\end{equation}
\begin{equation}
    j_{31}= \frac{\delta_1\left(B_2 g_2 d x_2\right)}{g_2 d x_1 d x_2}
\label{simple J31}
\end{equation}
\begin{equation}
    j_{32}=\frac{\delta_2\left(B_1 d x_1\right)}{g_2 d x_1 d x_2} 
\label{simple J32}
\end{equation}
which are the actual current density components computed using the simple finite difference (see the definition of $\delta_m$ given by equation [\ref{simple_diff}]) of the magnetic field components times the geometric factors in the adjacent cells. Again, we note that because of the monotonicity-preserving reconstructions used, the above $q$ factors are all between $0$ and $1$, and for smooth magnetic field profiles, the factors (raised to the power of $l=4$ in the code) are significantly less than 1 and therefore significantly reduce the magnitude of the first terms in the above numerical diffusion coefficients given in equations (\ref{eta12})-(\ref{eta32}).

When the high-order reconstruction (PDM) is used to compute the limitted current density components in equations (\ref{limited J12})-(\ref{limited J32}), then the diffusion coefficients $\eta_{12}$, $\eta_{13}$, $\eta_{23}$, $\eta_{21}$, $\eta_{31}$, $\eta_{32}$ are given by:
\begin{equation}
\eta_{12} = \frac{1}{2} \min \left( v_a, c \right) \, g_2 dx_2 + \frac{1}{2} \left| v_2\right| g_2 dx_2
\label{eta12_pdm}
\end{equation}
\begin{equation}
\eta_{13}= \frac{1}{2} \min \left( v_a, c \right) \, g_{31} g_{32} dx_3 + \frac{1}{2} \left| v_3\right| g_{31} g_{32} dx_3
\label{eta13_pdm}
\end{equation}
\begin{equation}
\eta_{23}= \frac{1}{2} \min \left( v_a, c \right) \, g_{31} g_{32} dx_3 + \frac{1}{2} \left| v_3\right| g_{31} g_{32} dx_3
\label{eta23_pdm}
\end{equation}
\begin{equation}
\eta_{21}= \frac{1}{2} \min \left( v_a, c \right)\, dx_1 + \frac{1}{2} \left| v_1\right| dx_1
\label{eta21_pdm}
\end{equation}
\begin{equation}
\eta_{31}= \frac{1}{2} \min \left( v_a, c \right) \, dx_1 + \frac{1}{2} \left| v_1 \right| dx_1
\label{eta31_pdm}
\end{equation}
\begin{equation}
\eta_{32}= \frac{1}{2} \min \left( v_a, c \right) \, g_2 dx_2 + \frac{1}{2} \left| v_2\right| g_2 dx_2,
\label{eta32_pdm}
\end{equation}
where the local Alfv\'{e}n speed $v_a$ is directly applied for the first terms in the above diffusion coefficients. In the above $c$ denotes the (reduced) speed of light used for the Boris correction.

\subsection{Non-adiabatic effects \label{sec:non-adiabatic}}

The internal energy equation (\ref{energy_eq}) takes into account the following non-adiabatic effects as listed in the right hand side of equation (\ref{non-adibatic-terms}): $\nabla \cdot \mathbf{q}_e$ the electron heat conduction, $Q_{\mathrm{rad}}$ the optically thin radiative cooling,  and $H_{\mathrm{em}}$ the empirical coronal heating. Furthermore, the effective heating resulting from the numerical diffusion of velocity and magnetic fields in the momentum and induction
 equations are evaluated and explicitly added to the internal energy in equation (\ref{energy_eq}) as $H_{\mathrm{num}}$. In the following we described the calculation of each of the non-adiabatic terms.
 
\subsubsection{Electron heat conduction}
 $\nabla \cdot \mathbf{q}_e$ is the electron heat conduction, where $\mathbf{q}_e$ denotes the electron heat flux. We use the form of $q_e$ as that given in \citet{vanderHolst:etal:2014}:
\begin{equation}
    {\bf q}_e = f_e {\bf q}_s+ (1-f_e) {\bf q}_H ,
\end{equation}
which combines the collisional heat flux denoted by ${\bf q}_s$ and the collisionless form denoted by ${\bf q}_H$ with an $r$ dependent weighting function:
\begin{equation}
    f_e = \frac{1}{1+(r/r_H)^2} ,
\end{equation}
where $r_H = 5 R_{\odot}$ with $R_{\odot}$ being the solar radius, such that the heat flux transition smoothly from the collisional form in the lower solar corona, to the collisionless form at large distances. 

For computing the collisional ${\bf q}_s$, we have used the hyperbolic heat conduction approach \citep{rempel2016extension} with the heat flux $\mathbf{q}_s$ given by:

\begin{equation}
    \frac{\partial \mathbf{q}_s}{\partial t}=\frac{1}{\tau_s} \left(-\kappa_0 T^{5 / 2} \hat{\mathbf{b}} \hat{\mathbf{b}} \cdot \nabla T-\mathbf{q}_s\right) ,
\label{eq:hbcond}
\end{equation}
where the first term in the right hand side is the collisional formulation of the electron heat flux of Spitzer with $\kappa_0 = 10^{-6} {\rm erg} \, {\rm s}^{-1} {\rm cm}^{-1} {\rm K}^{-7/2} $, and $\tau_s$ represents a finite time scale for $\mathbf{q}_s$ to evolve towards the Spitzer heat flux, which is set to

\begin{equation}
    \tau_s=\frac{\kappa_0 T^{7 / 2}}{e} \frac{(\Delta t)^2}{\left(f_{\mathrm{CFL}} \delta x_{\min }\right)^2},
\label{eq:tau_s}
\end{equation}
where $\Delta t$ denotes the dynamic Courant–Friedrichs–Lewy (CFL) time step (eq. [\ref{eq:CFL}]) due to all of the other advection terms, $f_{\mathrm{CFL}}$ is the CFL number used, and $\delta x_{\min} \equiv \min\!\left(dx_1,\, dx_2\, g_2,\, dx_3\, g_{31}\, g_{32}\right)$ is the minimum grid spacing. We use second-order finite difference to evaluate the gradient in equation (\ref{eq:hbcond}) at the appropriate cell interfaces. For computing the unit vector ${\hat{\bf b}}$ in equation (\ref{eq:hbcond}), we compute $b_1 = B_1/(B + \epsilon)$, $b_2=B_2/(B+ \epsilon)$, and $b_3 = B_3/(B+\epsilon)$, where $\epsilon$ is a tiny value ($10^{-99}$ used in the code), to avoid overflow when $B=0$.  Thus at the magnetic null point, ${\hat{\bf b}=0}$, and effectively the heat flux is set to zero.

The magnitude of ${\bf q}_s$ evaluated from equation (\ref{eq:hbcond}) is further limited by the saturation electron heat flux \citep{Fisher:etal:1985}:
\begin{equation}
    f_{\rm sat}=n_e (kT)^{3/2}/(4 \sqrt{m_e})=\frac{1}{8 \sqrt{2 \gamma}} \sqrt{\frac{m_p}{m_e}} \, (\gamma-1) \, e \, c_s
\end{equation}
where $n_e$ is the electron number density, $m_e$ and $m_p$ are respectivly the electron and proton masses, $k$ is the Boltzmann constant, $c_s$ is the sound speed. If the magnitude of ${\bf q}_s$ evaluated from equation (\ref{eq:hbcond}) goes above $f_{\rm sat}$, it is set to $f_{\rm sat}$.

Implementing the above hyperbolic approach for the collisional electron heat conduction eliminates the stringent time step constraint imposed by the traditional parabolic form of heat conduction as a diffusive process, which under typical coronal conditions can be orders of magnitude more severe compared to the dynamic CFL time step determined from the fast mode speed \citep[e.g][]{rempel2016extension}. As shown by \citet{rempel2016extension} (Section 3.6), under the limit of the saturation electron heat flux where the Spitzer conductivity is appropriate, the hyperbolic approach gives an accurate description of the heat conduction process. In our practical simulations with the MFE code using the above hyperbolic formulation (eqs. [\ref{eq:hbcond}] [\ref{eq:tau_s}]), we found that the actual ${\bf q}_s$ obtained during the simulation closely tracks the Spitzer flux $\kappa_0 T^{5 / 2} \hat{\mathbf{b}} \hat{\mathbf{b}} \cdot \nabla T$ to a high accuracy (with a relative difference on the order of $10^{-4}$.)

At large heliocentric distances where the plasma becomes sufficiently rarefied and the electron mean free path increases, the electron heat flux transitions to the collisionless form ${\bf q}_H$, estimated as \citep{Hollweg:1976}:
\begin{equation}
    {\bf q}_H = \frac{3}{4} \alpha p {\bf v}
\end{equation}
where $\alpha$ is an unknown parameter of order unity. Here we set the value of $\alpha=1.05$ as used in \citet{vanderHolst:etal:2014}. The collisionless electron heat conduction effectively enhances the enthalpy flux of the solar wind flow at large distances.

\subsubsection{Optically Thin Radiative Cooling}
$Q_{\mathrm{rad}}$ in equation (\ref{non-adibatic-terms}) is the optically thin radiative cooling:

\begin{equation}
    Q_{\mathrm{rad}}=N^2 \Lambda(T),
\end{equation}
where $N=\rho / m_p$ is the proton number density assuming a fully ionized hydrogen gas, with $m_p$ being the proton mass, and the radiative loss function $\Lambda(T)$ implemented in the code is the analytical expression of $f_n (T)$ given by equations (25) and (26) in \citet{athay1986radiation}, who examined the effects of optical thickness in the computation of the radiative loss rates for the solar atmosphere.

\subsubsection{Empirical Coronal Heating \label{sec:coronal_heating}}
$H_{\mathrm{em}}$ in equation (\ref{energy_eq}) is an empirical coronal heating which is typically a function of $r$ and can be specified based on the problems of interest. For example in the simulation described below of solar wind with a dipole magnetic field (section \ref{sec:solarwind}), we have used the following empirical coronal heating function:
\begin{equation}
    H_{\mathrm{em}}=\frac{F}{L_H} \exp \left[-\left(r-R_{\odot} \right) / L_H\right]
\label{eq:coronal_heating}
\end{equation}
with $F=3.1 \times 10^5 \mathrm{erg} / \mathrm{s} / \mathrm{cm}^2, L_H= 0.77 R_{\odot}$. This is the empirical coronal heating function used in the 1-D solar wind model of \citet{Withbroe:1988} for modeling the solar polar coronal hole during solar minimum.

\subsubsection{Numerical Viscous and Ohmic Heating}
We also evaluate the numerical viscous and Ohmic heating due to the numerical diffusion and add their contribution to the internal energy equation as the term $H_{\rm num}$:
\begin{equation}
    H_{\text {num }}={\cal S} : \nabla {\bf w} + {\bf E}^* \cdot {\bf j},
\label{H_num}
\end{equation}
where the first term on the RHS is the numerical viscous heating and second term is the numerical resistive heating. ${\cal S}$ is a numerical viscous stress tensor whose elements consist of those diffusive upwind momentum fluxes given by the last terms of equations (\ref{m1 flux1})-(\ref{m3 flux3}). Specifically, the last terms of equations (\ref{m1 flux1}) (\ref{m1 flux2}) (\ref{m1 flux3}) multiplied by $-1$ comprise the first-row elements of the ${\cal S}$ matrix,  the last terms of equations (\ref{m2 flux1}) (\ref{m2 flux2}) (\ref{m2 flux3}) multiplied by $-g_2^2$ comprise the 2nd-row elements of ${\cal S}$, and the last terms of equations (\ref{m3 flux1}), (\ref{m3 flux2}), (\ref{m3 flux3}) multiplied by $-(g_{31}g_{32})^2$ comprise the 3rd-row elements of ${\cal S}$. The vector ${\bf w}=(v_1, \, v_2/g_2, \, v_3/g_{31}g_{32})$. The inner product of the matrices ${\cal S}$ and $\nabla {\bf w}$ gives the viscous heating. Here the gradient of ${\bf w}$ is evaluated using second-order finite difference. For the resistive heating term in equation (\ref{H_num}), ${\bf E}^*$ is the diffusive electric field given by equations (\ref{diffemf1}) (\ref{diffemf2}) (\ref{diffemf3}), and vector ${\bf j}$ is the electric current density given by ${\bf j} = (j_{12}-j_{13}, \, j_{23}-j_{21}, \, j_{31}-j_{32})$, whose components are given in equations (\ref{simple J12}), (\ref{simple J13}), (\ref{simple J21}), (\ref{simple J23}), (\ref{simple J31}), (\ref{simple J32}). Another similar approach to account for heating due to numerical diffusion in low plasma-$\beta$ simulations is employed in the MURaM code \citep{rempel2016extension} which solves the plasma energy (internal energy plus kinetic energy) equation instead of the internal energy equation. By solving the plasma energy equation, the (numerical) viscous heating is implicitly included while the resistive heating due to numerical diffusion is similarly evaluated and added to the plasma energy equation explicitly \citep[see section 2.6 of][]{rempel2016extension}.

\subsection{Time Integrator}
\label{sec:time_stepping}

To advance the solution in time, we adopt the low‐storage third‐order Runge–Kutta (RK3) method \citep{williamson1980low}. Write the system of equations (\ref{discretized rho}), (\ref{discretized energy}), (\ref{eq:mom1}), (\ref{eq:mom2}), (\ref{eq:mom3}), (\ref{eq:B1}), (\ref{eq:B2}), (\ref{eq:B3}), and (\ref{eq:hbcond}) in a general form as
\begin{equation}
    \frac{\partial \boldsymbol{u}}{\partial t} = \boldsymbol{L}(\boldsymbol{u}, t),
\end{equation}
where $\boldsymbol{u}$ represents the state variables and $\boldsymbol{L}$ is a function of $\boldsymbol{u}$ and time. The RK3 formulation is given by
\begin{align}
  &\boldsymbol{u}^{(0)} = \boldsymbol{u}^n, \notag \\
  &\boldsymbol{k}^{(i)} = \alpha_i\, \boldsymbol{k}^{(i-1)} +  \boldsymbol{L}\Bigl(\boldsymbol{u}^{(i-1)}, t^n + \gamma_i\, \Delta t\Bigr), \quad \text{for } i \in \{1, 2, 3\}, \notag \\
  &\boldsymbol{u}^{(i)} = \boldsymbol{u}^{(i-1)} + \beta_i\, \Delta t\, \boldsymbol{k}^{(i)}, \quad \text{for } i \in \{1, 2, 3\}, \notag \\
  &\boldsymbol{u}^{n+1} = \boldsymbol{u}^{(3)}.
\end{align}
The coefficients $\alpha_i$, $\beta_i$, and $\gamma_i$ employed in the scheme are provided in Table~\ref{tab:rk3_label}.
\begin{deluxetable}{cccc}[htb!]  \tablenum{1}  \tablecaption{Constants for the low-storage RK3 method \label{tab:rk3_label}}
\tablewidth{0pt}
\tablehead{
    \colhead{i} & \colhead{$\alpha_i$} & \colhead{$\beta_i$} & \colhead{$\gamma_i$}
}
\startdata
1 & 0 & $1/3$ & 0 \\
2 & $-5/9$ & $15/16$ & $1/3$ \\
3 & $-153/128$ & $8/15$ & $3/4$ \\
\enddata
\end{deluxetable}
The dynamic CFL time step $\Delta t$ is determined by
\begin{equation}
\Delta t = f_{\mathrm{CFL}}\, \min\!\left(\frac{\delta x_{\min}}{\left|v\right|+c_f} \right),
\label{eq:CFL}
\end{equation}
where $v$ represents the cell-centered fluid velocity and $c_f$ is the cell-centered fast-mode speed, for which the Alfv\'{e}n speed is given by the minimum of the Alfv\'{e}n speed $v_a$ and the reduced speed of light $c$. The minimum grid spacing is defined as $\delta x_{\min} \equiv \min\!\left(dx_1,\, dx_2\, g_2,\, dx_3\, g_{31}\, g_{32}\right)$. In our simulations, the CFL number $f_{\mathrm{CFL}}$ is typically set to 0.25.

\subsection{Yin-Yang Grid for Full Sphere Problem}
\begin{figure}[ht!]
\centering
\includegraphics[width=0.6\textwidth]{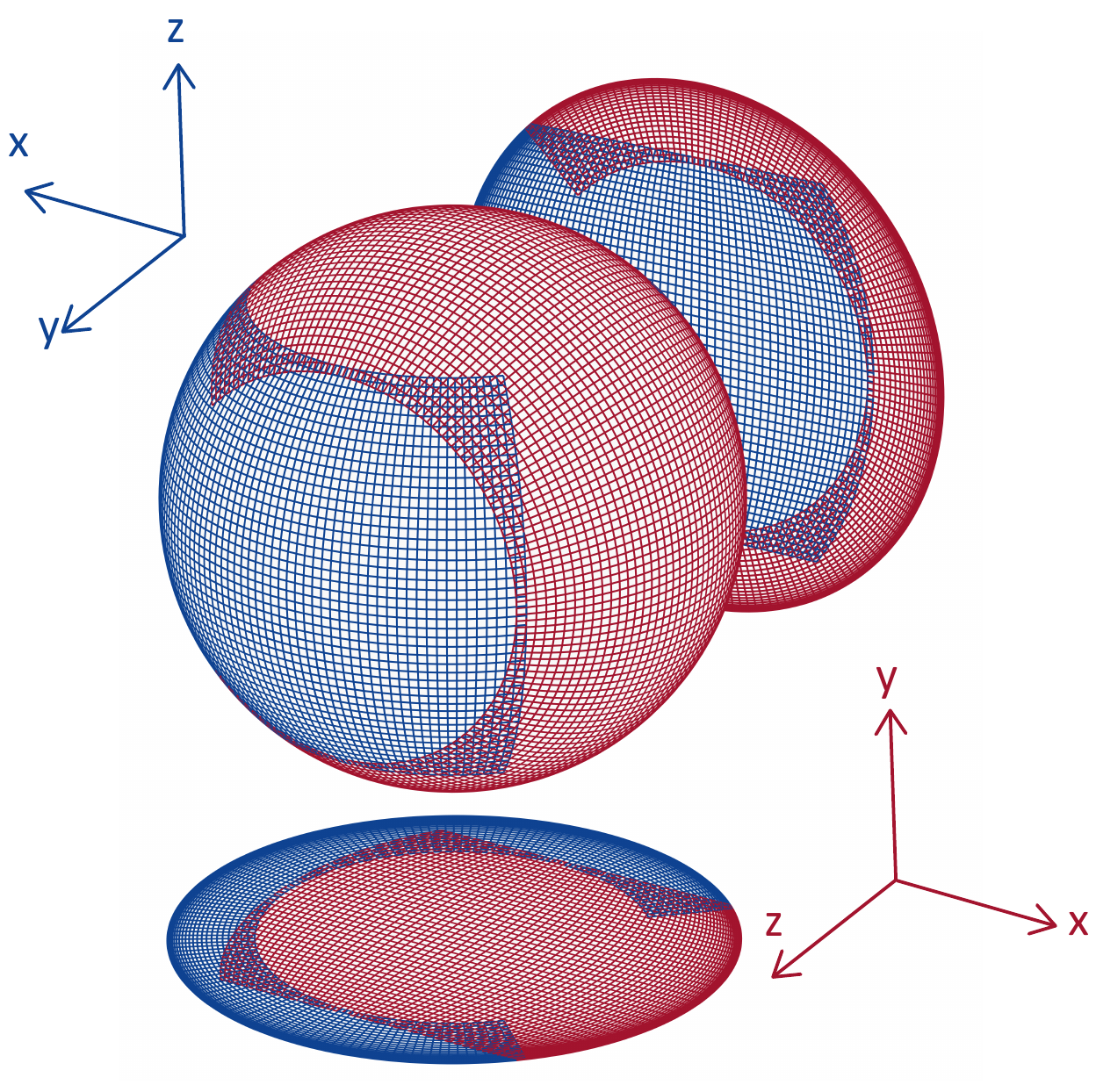}
\caption{A schematic of the Yin-Yang grid: two identical sized overlapping partial spherical shell domains with $r$-$\theta$-$\phi$ grids (blue grid representative of Yin and red grid representative of Yang) with orthogonal polar axes, that cover the whole sphere.
\label{fig:yin-yang grid}}
\end{figure}

Full‐sphere simulations using conventional spherical meshes typically face the following main challenges: grid anisotropy and singularity near the axis, and stringent limitations on time‐step sizes. To mitigate these issues, we implement the Yin–Yang overset grid introduced by \citet{kageyama2004yin}. The Yin–Yang grid consists of two identical sized overlapping partial spherical shell domains with $r$-$\theta$-$\phi$ grids with orthogonal polar axes, that cover the whole sphere (see the schematic in Figure \ref{fig:yin-yang grid}).
Each of the partial spherical shell domain (in its own spherical polar coordinate system) is given by $(\pi / 4 - \delta \leq \theta \leq 3 \pi / 4 +\delta ) \cap(-3 \pi / 4 -\delta \leq \phi \leq 3 \pi / 4 +\delta)$, where $\delta $ denotes a small overlapping region which is here set to be slightly greater than 0.5 of the larger cell size of the Yin- and Yang-grids. In \citet{kageyama2004yin} the transformation between the Yin and Yang coordinate systems of the position coordinates and the scalar quantities are given in Eqs. (2)-(9) and vector quantities in Eqs. (22)-(25).

Within each of the Yin- and Yang-grids, the code solves the single fluid semi-relativistic MHD equations in its own spherical $r$-$\theta$-$\phi$ grid domain as described in the previous sections. The necessary ghost zone values of the primitive variables for the Yin- and Yang-grids are obtained by bilinear interpolation into the overlapping interior zones of the adjacent Yin- and Yang-domains. Electric fields in the ghost zones are calculated from the interpolated primitive variables.
The code is parallelized with MPI using 3D domain decomposition. In each of the Yin- and Yang-domains, the MPI 3D domain decomposition is carried out in a way such that each individual sub-domain's grid size is identical across the Yin- and Yang-domains to ensure load balance. The implementation of the Yin-Yang grid essentially boils down to finding an efficient way for updating the ghost zones of the Yin- and Yang-domains via MPI communications between the processors adjacent to the Yin-Yang boundaries. We have implemented this using MPI non-blocking point-to-point communication calls between the boundary processors. At the beginning, the code computes once and stores the interpolation coefficients needed for computing the variables in the ghost zones using bilinear interpolation into the overlapping interior zones in the adjacent Yin- or Yang-domain.  Also at the beginning, the code groups the ghost zones into messages in terms of their senders, receivers, sizes, and data buffer locations so that each MPI process is given the information on the MPI sends and receives it needs to perform to update the Yin- and Yang-domain’s ghost zones at each time step.
As a result, our implementation of the Yin-Yang grid code achieves efficient Yin-Yang domain boundary communication, such that the code
shows good scaling of speed up with the number of processors for a typical 3D global corona simulation such as the solar wind simulation described in section \ref{sec:solarwind}.

\section{Test Problems}\label{sec:tests}

\subsection{Nonlinear Circularly Polarized Alfv\'{e}n waves}
We employ the two-dimensional circularly polarized nonlinear Alfv\'{e}n wave test described in \citet{toth2000b}, which is an exact nonlinear MHD solution, to demonstrate the accuracy of the numerical scheme in the nonlinear regime. The computational domain covers $x \in [0,1/\cos\alpha]$ and $y \in [0,1/\sin\alpha]$, where $\alpha = \pi/3$ is the wave propagation angle relative to the $x$-axis. Because the numerical fluxes differ in the $x$- and $y$-directions, this setup is genuinely multi-dimensional. We apply periodic boundary conditions in both directions. The initial conditions are $\rho = 1$, $p = 0.1$, $v_{\perp} = (0.1/\sqrt{4\pi})\sin(2\pi x_{\|})$, $B_{\perp} = 0.1\sin(2\pi x_{\|})$ and $v_{z} = (0.1/\sqrt{4\pi}) \cos(2\pi x_{\|})$, $B_{z}= 0.1\cos(2\pi x_{\|})$, with $\gamma = 5/3$. Here $x_{\|} = x \cos\alpha + y \sin\alpha$ defines the coordinate along the wave vector (propagation) direction. The quantities $u_{\perp}$ and $B_{\perp}$ denote the velocity and magnetic field components perpendicular to the wave vector. The parallel and perpendicular magnetic field components are related to the Cartesian components by $B_{\perp} = B_{y}\cos\alpha - B_{x}\sin\alpha$ and $B_{\|} = B_{x}\cos\alpha + B_{y}\sin\alpha$. The initial parallel field strength is set to $B_{\|0} = 1$. To ensure a divergence-free initial field, we initialize the magnetic field through a vector potential with components:
\begin{equation}
    A_x=B_{\| 0} \sin \alpha z
\end{equation}
\begin{equation}
    A_y=-B_{\| 0} \cos \alpha z+\frac{0.1}{2 \pi \cos \alpha}
\sin [2 \pi(x \cos \alpha+y \sin \alpha)]
\end{equation}
\begin{equation}
    A_z= \\
\frac{0.1}{2 \pi} \cos [2 \pi(x \cos \alpha+y \sin \alpha)].
\end{equation}

\begin{figure}[htb!]
\centering
\includegraphics[width=0.6\textwidth]{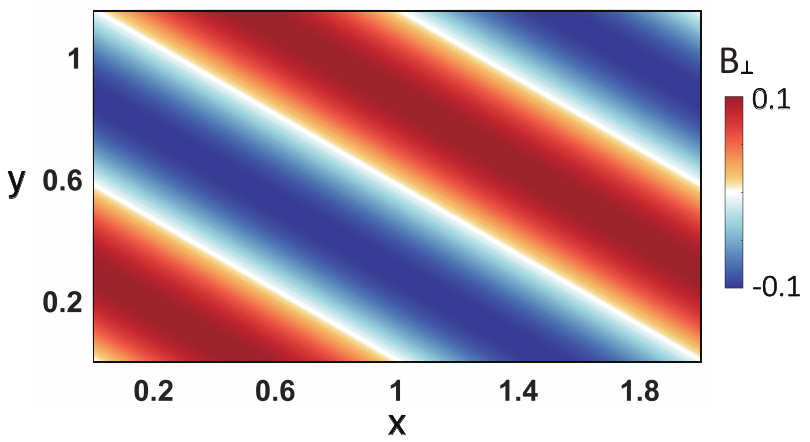}
\caption{The spatial distribution of $B_{\perp}$ of nonlinearly circularly polarized Alfv\'{e}n waves  after propagating one
 wavelength, using the PLM and a grid of $256 \times 256 $ cells .}
\label{fig:NLA_bperp}
\end{figure}

Figure~\ref{fig:NLA_bperp} shows the spatial distribution of $B_{\perp}$ at $t = 1.0$ (one orbit) on a $256 \times 256$ grid using the PLM. The wave’s amplitude is well preserved, and the wavefronts remain flat, confirming the numerical method's fidelity.

\begin{figure}[htb!]
\centering
\includegraphics[width=0.45\textwidth]{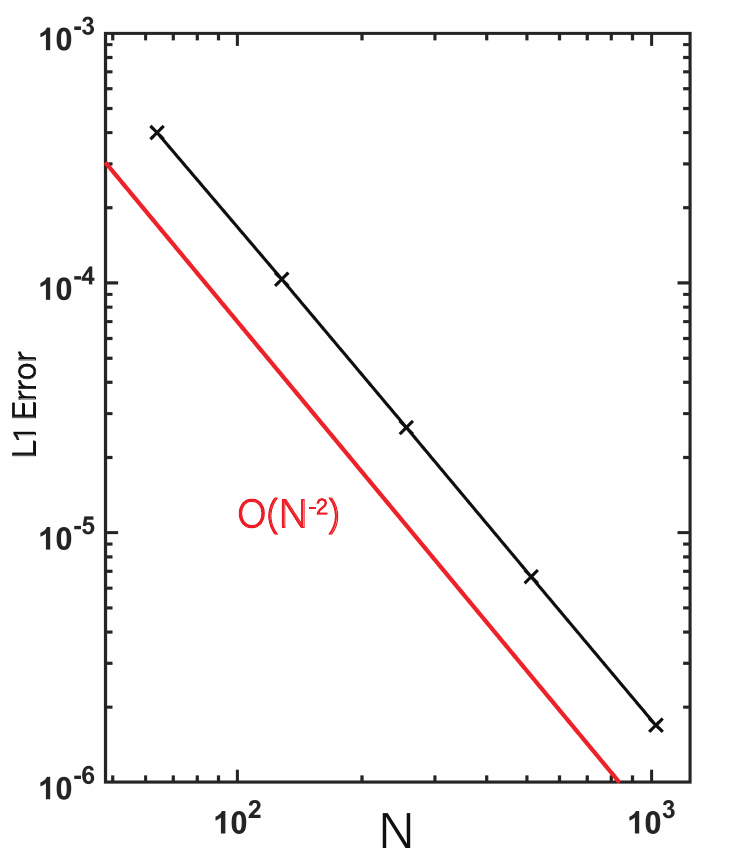}
\caption{Convergence of the nonlinear polarized Alfv\'{e}n waves test}
\label{fig:NLA_l1}
\end{figure}

To verify the numerical convergence, we perform simulations of this smooth nonlinear problem using increasingly finer grids ($N\times N$ cells, where $N = 64,128,256,512,1024$). Figure~\ref{fig:NLA_l1} illustrates the mean $L1$-norm errors of all primitive variables as a function of grid resolution, demonstrating clear second-order convergence.

\subsection{Brio-Wu Shock Tube}

\begin{figure}[ht!]
\centering
\includegraphics[width=0.85\textwidth]{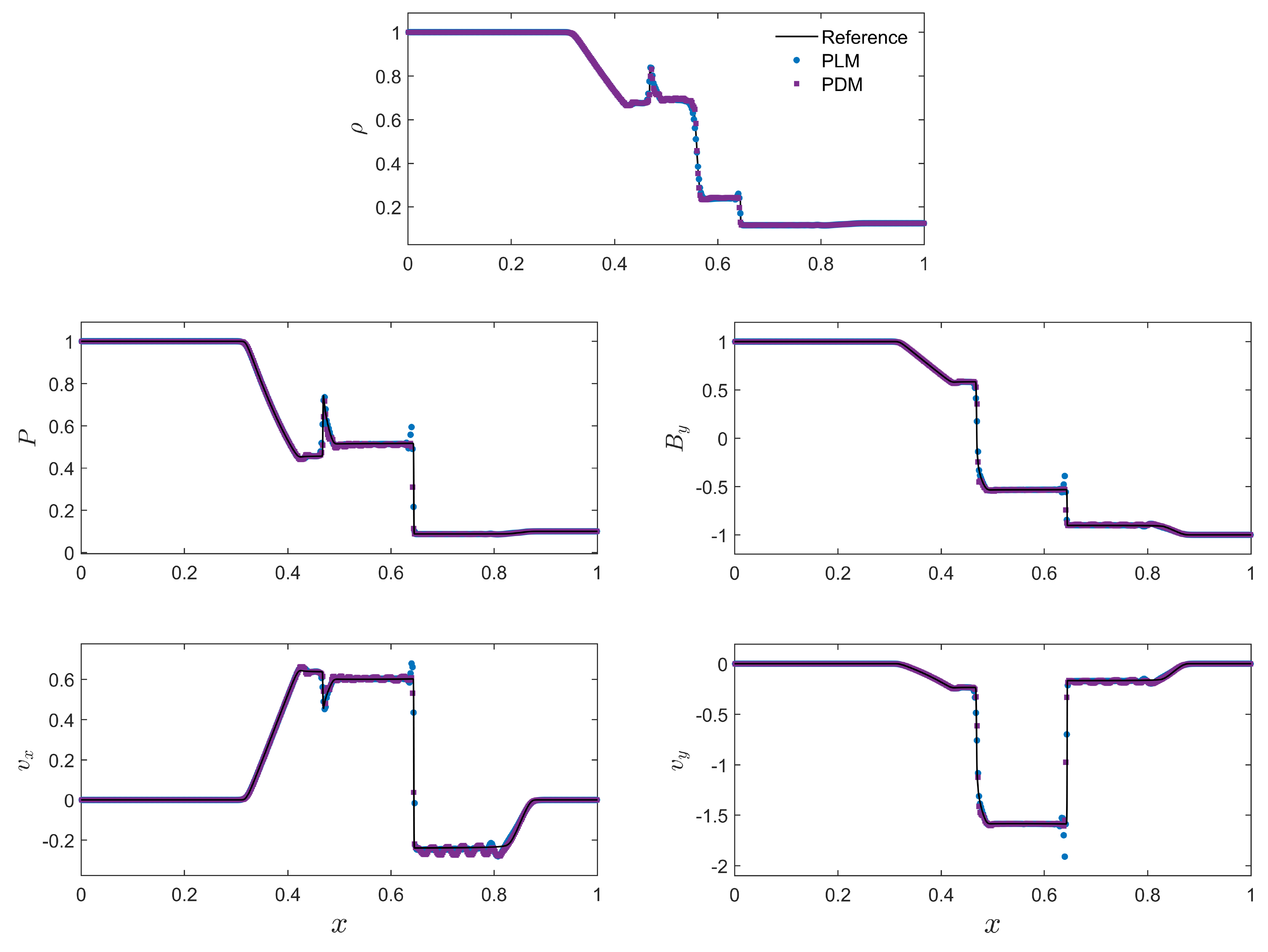}
\caption{Brio-Wu shocktube problem using different reconstructions and $N=512$ grid cells, data taken at $t=0.1$. The reference result (black solid lines) is obtained using a 1D conservative code. 
\label{fig:WB shocktube}}
\end{figure}

The one-dimensional Brio–Wu shock tube \citep{brio1988upwind} is used here to evaluate the numerical scheme on nonlinear problems involving strong shocks. The computational domain spans $x \in [0,1]$ with $N_x = 512$ grid cells. The initial conditions are:
\begin{equation}
    \left(\rho, B_{x}, B_{y},  P, \right)= \begin{cases}(1.0,0.75,1.0,1.0) & (x < 0.5) \\ (0.125,0.75,-1.0,0.1) & (x\geq0.5)\end{cases}
\end{equation}
and $B_z=0$, $v_x = v_y = v_z =0$, and with $\gamma =2$. Figure~\ref{fig:WB shocktube} compares the simulation results of using the PLM (blue dots) and the PDM (purple dots) at $t = 0.1$ with a high-resolution ($N_x=16384$) reference solution satisfying the Rankine-Hugoniot conditions, obtained using a first-order conservative 1D code with total energy equation (black solid lines). The shocks and characteristic features are well captured by both the PLM and PDM cases, although some significant spikes are observed at the shock front in the PLM case, indicating an insufficient numerical diffusion with the use of high $l=4$ with the PLM in this case. The PDM case on the other hand shows excellent agreement with the reference solution, which demonstrates its accuracy and robustness in handling nonlinear MHD shocks.

\subsection{Orszag-Tang Vortex}

We run the classical two-dimensional Orszag–Tang Vortex simulation \citep{orszag1979small} to test the effectiveness of the numerical schemes on handling shock formations and interactions. The test simulation is done within a square domain of $x \in [0,1]$, $y \in [0,1]$ with periodic boundary condition in both directions. The initial density and pressure are uniform within the simulation domain:  $\rho=25/36\pi$, $p=5/12\pi$. The initial velocities are set as $v_{x}=-\sin (2 \pi y)$ and $v_{y}=\sin (2 \pi x)$ and the initial magnetic field is initialized via the vector potential $A_{z}=B_{0} \left(\cos (4 \pi x)/ 4\pi + \cos (2 \pi y)/ 2\pi \right)$  with $B_{0} = 1$.

\begin{figure}[ht!]
\plotone{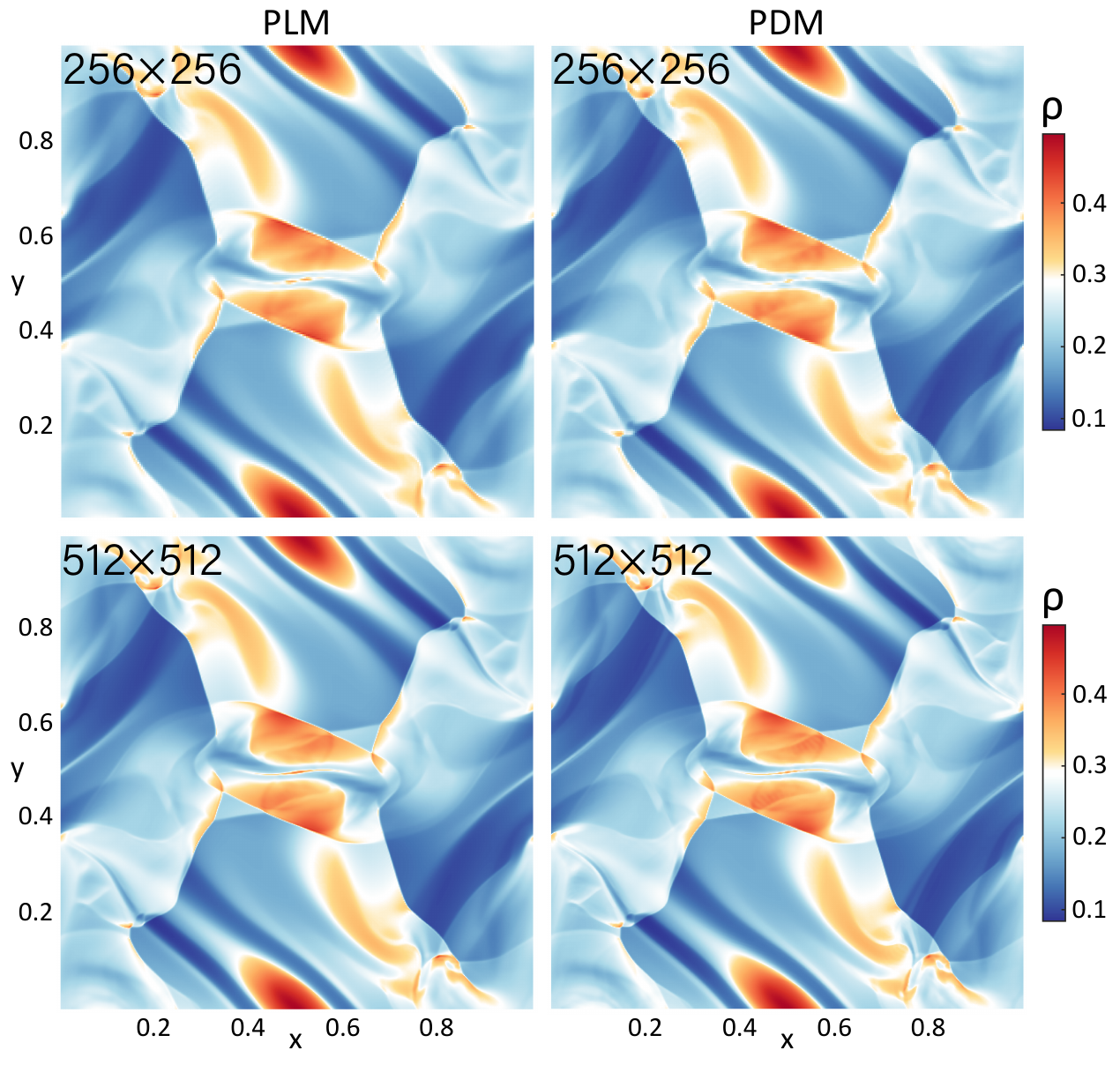}
\caption{The spatial distribution of plasma density of Orszag-Tang simulations at $t=0.5$ using different reconstructions and resolutions: $256 \times 256 $ (top) and $512 \times 512 $ (bottom), piecewise linear method (left) and partial donor cell method (right).
\label{fig:Ot_05_rho}}
\end{figure}

\begin{figure}[ht!]
\plotone{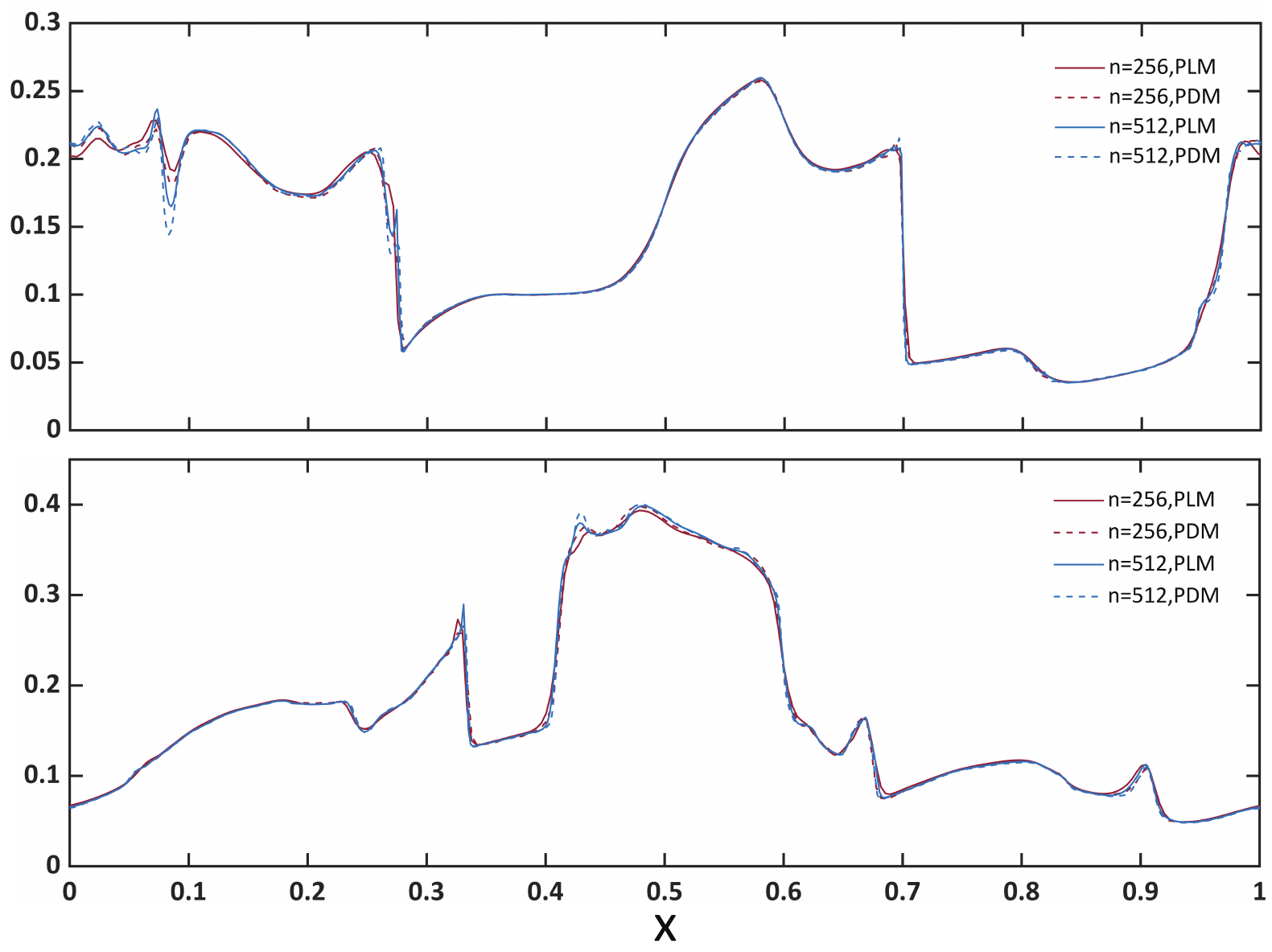}
\caption{The line profiles of plasma pressure at $y = 0.3125$ (top) and $y=0.4277 $(bottom) from four Orszag–Tang simulations at $t=0.5$. The results are shown for  $256\times256$ grid (red line) and $512\times512$ grid (blue line), PLM reconstruction (solid line) and PDM reconstruction (dashed line).
\label{fig:Ot_05_profile}}
\end{figure}

\begin{figure}[ht!]
\plotone{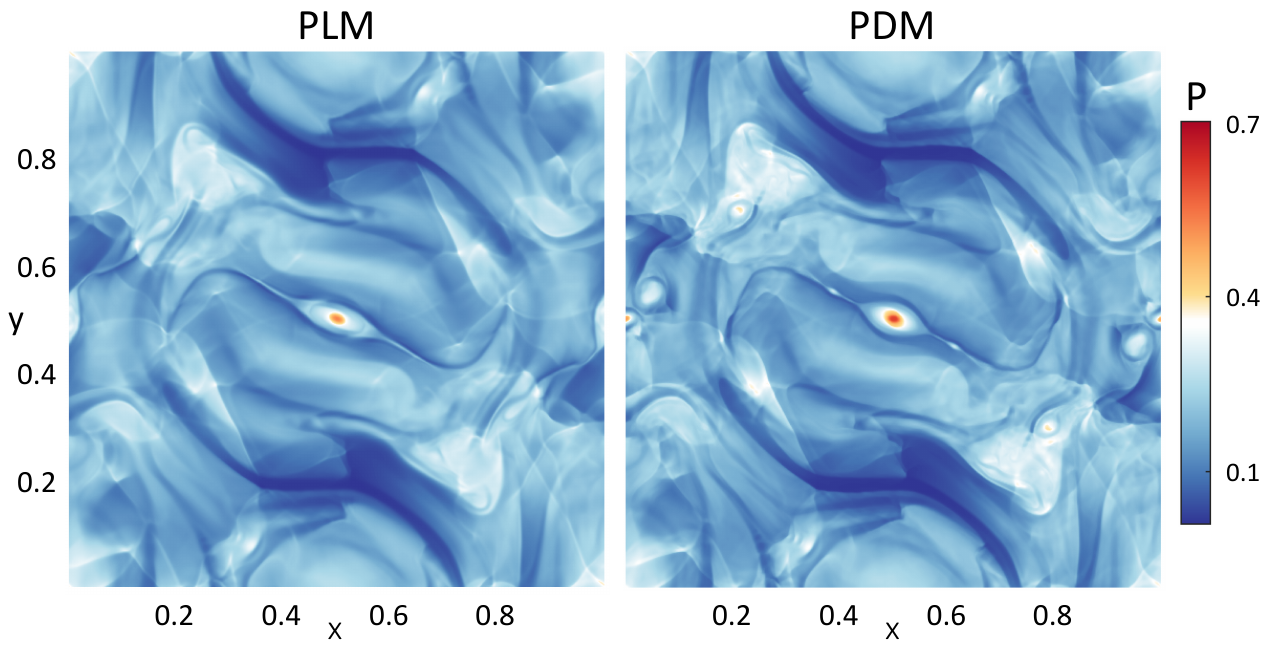}
\caption{The spatial distribution of plasma pressure of Orszag-Tang simulations at $t=1$ using a $512 \times 512$ grid: results obtained with PLM (left) and PDM (right).
\label{fig:Ot_10_p}}
\end{figure}

Figure \ref{fig:Ot_05_rho} shows the spatial distribution of density at $t=0.5$ for different reconstruction methods (PLM and PDM) and grid resolutions ($256^2$ and $512^2$). The results preserve symmetry nicely and no spurious oscillation is observed. A more detailed comparison is given in Figure \ref{fig:Ot_05_profile}, showing the line profile of pressure at $t=0.5$ along $y=0.3125$ and $y=0.4277$. The results can be directly compared with other codes, e.g., \citet{londrillo2000high,stone2008athena}, demonstrating the capability of the numerical methods to accurately capture shocks while maintaining non-oscillatory solutions. At later times, complex shock–vortex interactions give rise to an inclined, twisted current sheet at the center, where a central magnetic island is observed. This feature is indicative of low numerical diffusion, which favors the development of the tearing mode instability \citep{landi2015resistive,berta20244th}. Figure~\ref{fig:Ot_10_p} displays the pressure distribution at $t=1$, showing that both the PLM and PDM algorithms accurately capture the central magnetic island and the rich fine-scale structures.

\subsection{Field-loop Advection}

\begin{figure}[ht!]
\plotone{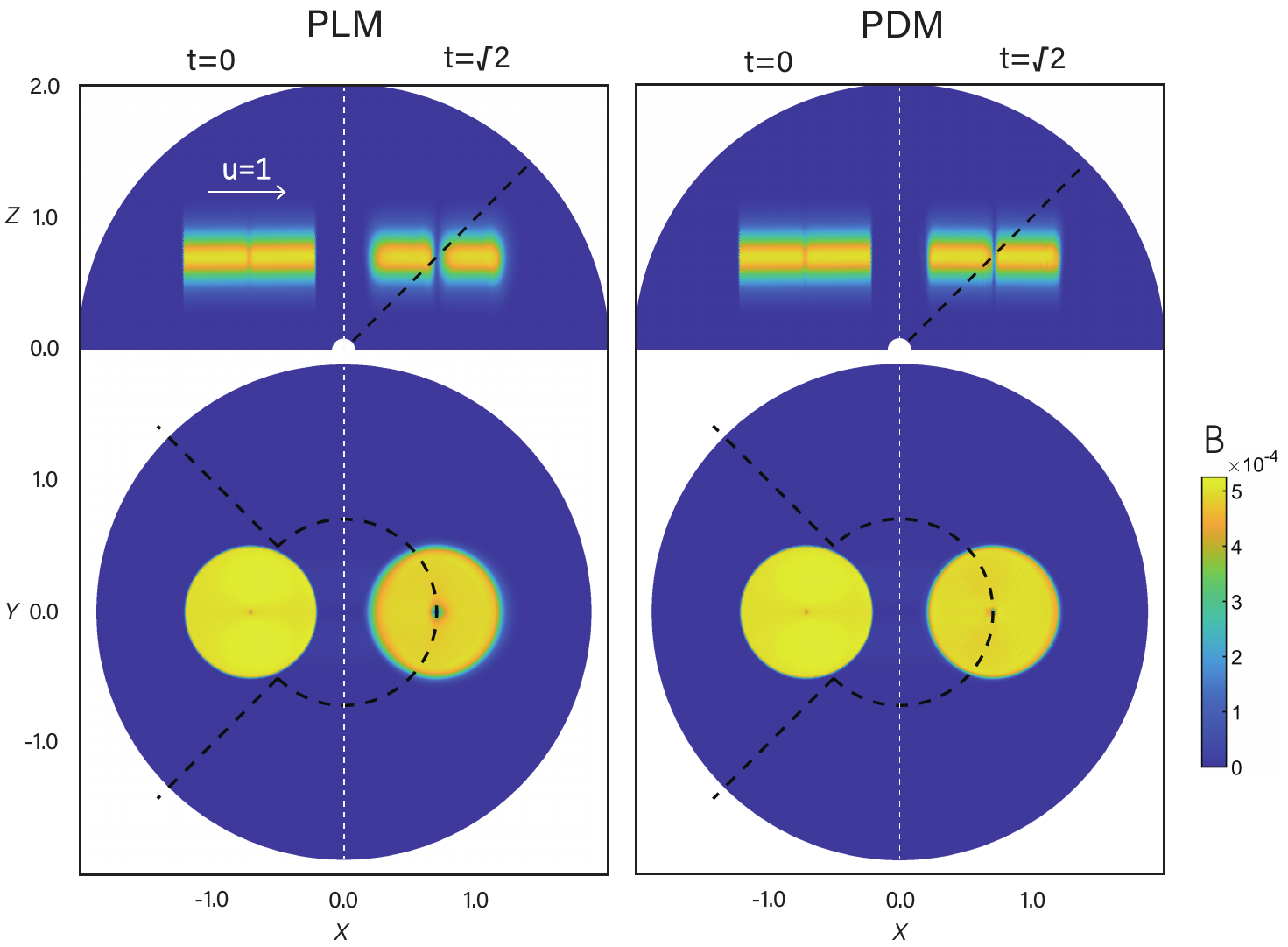}
\caption{Magnetic field in three-dimensional magnetic field loop advection test  in spherical coordinate at $t=\sqrt{2}$. The top panels show vertical slices through the $y = 0$ plane, and the bottom panels show horizontal slices through the
$z = \sqrt{2}/ 2$ plane. The black dashed lines represent the Yin-Yang boundaries. The left panel shows the results obtained with PLM reconstruction and the right panel shows the results obtained with PDM reconstruction.
\label{fig:field_loop}}
\end{figure}

The magnetic field loop test, first introduced in \citet{gardiner2005unsplit}, involves the advection of a weakly magnetized field loop in a constant velocity field. The plasma $\beta$ is set sufficiently high such that the loop behaves nearly as a passive scalar. This test thus represents an MHD analog to the multidimensional square-wave advection problem. Although conceptually simple, this test is numerically challenging; an induction equation algorithm that fails to properly enforce the divergence-free constraint and incorporate appropriate diffusion will yield a significantly distorted solution. We adopt the setup described in \citet{stone2020athena++}, in which a magnetic field loop is advected within a three-dimensional full sphere domain. The magnetic loop configuration is initialized with a vector potential:
\begin{equation}
\begin{aligned}
A_z= B_0 \exp \left[-\frac{\left(z-z_0\right)^2}{\sigma^2}\right]  \times \max \left(R-\sqrt{\left(x-x_0\right)^2+\left(y-y_0\right)^2}, 0\right),
\end{aligned}
\end{equation}
where $x_0 = -\sqrt{2}/2$, $y_0 = 0$, and $z_0 = \sqrt{2}/2$ denote the coordinates of the loop center, $B_0 = 5 \times 10^{-4}$ is the magnetic field strength, and $R = 0.5$ and $\sigma = 0.2$ represent the radius and thickness of the loop, respectively. The velocity field is a uniform flow in the $x$ direction with $v_x = 1$, and $\rho =1$, $p = 0.5$ everywhere. The computation is carried out in a full sphere with $0.1 \leq r \leq 2$ using a grid resolution of $240 (r)\times120(\theta)\times360(\phi)$ for each Yin–Yang component. Figure~\ref{fig:field_loop} shows the magnetic field solutions on constant-$z$ and constant-$y$ planes obtained with different reconstruction schemes. For both schemes, the loop is well preserved even after crossing the Yin–Yang boundaries (indicated by black dashed lines). Compared with the PLM scheme, the PDM method exhibits sharper edges, reflecting its low-diffusion characteristics. The three-dimensional magnetic field loop advection test in spherical coordinates demonstrates the accuracy and robustness of the numerical algorithms and the Yin–Yang implementation.

\section{Solar wind with a dipole magnetic field \label{sec:solarwind}}
In this section, we present a global solar wind simulation employing a simple dipole magnetic field configuration, representative of solar minimum conditions. We use a computational grid consisting of $840(r) \times 120 (\theta) \times 360 (\phi)$ for each Yin and Yang component, resulting in a uniform horizontal resolution of $0.75^{\circ}$. The radial extent of the domain is $[R_{\odot}, 10 R_{\odot}]$ with a uniform resolution of $7.46 \, {\rm Mm}$.

\begin{figure}[htb!]
\centering
\includegraphics[width=0.4\textwidth]{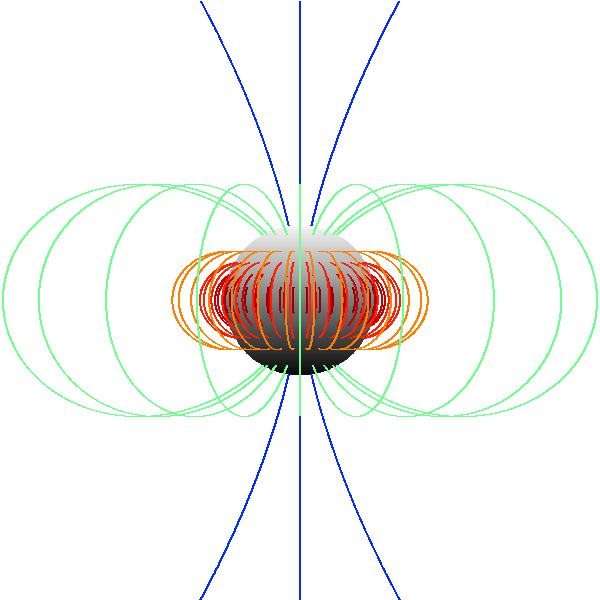}
\includegraphics[width=0.5\textwidth]{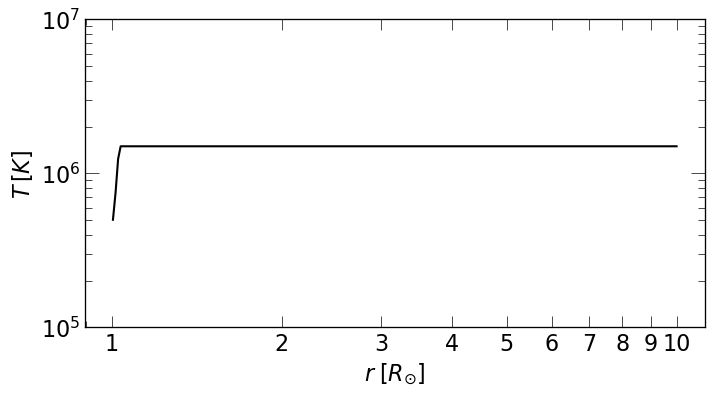}
\caption{Initial dipole magnetic field lines (left) and the initial temperature profile for the 1-D hydrostatic corona (right). The gray-scale on the sphere in left panel shows the dipolar normal magnetic field distribution $B_r$ on the solar surface}
\label{fig:solwind_inittemp}
\end{figure}

We initialize the simulation domain with a dipole magnetic field (left panel of Figure \ref{fig:solwind_inittemp}) given by:
\begin{equation}
{\bf B} = - \nabla \Phi,
\end{equation}
where
\begin{equation}
    \Phi = 5\, \frac{\cos(\theta)}{r^2},
\end{equation}
expressed in units of ${\rm G} R_{\odot}$. This yields a polar magnetic field strength of $10$ G. The initial plasma conditions are set using a one-dimensional hydrostatic atmosphere with a radial temperature profile shown in the right panel of Figure~\ref{fig:solwind_inittemp}.
The precise choice of the initial temperature profile does not significantly affect our results, since the corona and solar wind structure in the final steady-state are primarily shaped by the imposed coronal heating and magnetic field evolution. In this simulation, we adopt an empirical coronal heating function (eq. [\ref{eq:coronal_heating}],
section \ref{sec:coronal_heating}), originally used by \citet{Withbroe:1988} for modeling the solar-minimum coronal hole conditions. All other non-adiabatic processes, such as electron heat conduction, optically thin radiative losses, and numerical viscous and resistive heating, are handled as detailed in Section~\ref{sec:non-adiabatic}.

Here we describe the inner and outer boundary conditions used for the simulation. Let ``is" and ``ie" denote respectively the zone indices for the start and end of the 1-direction ($r$-direction) grid in the simulation domain, and hence ``is-1'' and ``is-2'' are the indices for the two ghost zones below the inner boundary surface, and ``ie+1", ``ie+2", ``ie+3" are the indices for the three ghost zones above the outer boundary surface. At the bottom first layer cell centers (at $x_{1b,is}$), we impose a fixed temperature of $T_{is} = 5 \times 10^5 {\rm K}$, and a time dependent base pressure $p_{is}$ according to:
\begin{equation}
\frac{ \partial p_{is}}{\partial t}
= \frac{1}{\tau} \left ( p_{\rm reb} - p_{is} \right )
\label{eq:lbpressure}
\end{equation}
where,
\begin{equation}
p_{\rm reb} = C \, f_c ,
\label{eq:p_reb}
\end{equation}
\begin{equation}
f_c = \kappa_0 T_{is+1/2}^{5/2} \left (\frac{dT}{dr} \right )_{is+1/2},
\label{eq:downward_fc}
\end{equation}
where $T_{is+1/2}$ and $(dT/dr)_{is+1/2}$ are respectively the average and centered finite difference of the $T$ values in the adjacent cell centers at $x_{1b,is}$ and $x_{ib,is+1}$, and the initial $p_{is}$ is simply set to the $p_{reb}$ given the initial temperature profile (right panel of Figure \ref{fig:solwind_inittemp}). The above formulation drives the base pressure $p_{is}$ towards a balance value $p_{\rm reb}$ proportional to the downward heat conduction flux $f_c$, as given by the radiative energy balance (reb) model of \citet{Withbroe:1988}, over a time scale of $\tau$, to represent the effect of chromospheric evaporation. For the simulation here we have used $C = 6.6 \times 10^{-7}$ in CGS units and $\tau = 1071 $ sec. The temperature and pressure (and density) in the inner ghost zones (at $x_{1b,is-1}$ and $x_{1b,is-2}$) are simply set to the values for hydrostatic equilibrium given $T_{is}$ and $p_{is}$ at $x_{1b,is}$.

For the magnetic field and the velocity field, we impose the ``line-tied'' inner boundary condition as follows. On the inner boundary surface at $x_{1a,is}=R_{\odot}$ we set the horizontal (2-, 3-directions) electric field and the vertical (1-direction) velocity field to zero: $E_2(x_{1a,is}, x_{2b,j}, x_{3a,k})=0$, $E_3(x_{1a,is}, x_{2a,j}, x_{3b,k})=0$, $v_1(x_{1a,is}, x_{2b,j}, x_{3b,k})=0$. We also set all of the electric field and velocity field components to zero in all of the inner ghost zones. Thus the magnetic field in the ghost zones and the normal magnetic field $B_1$ on the inner boundary surface (at $x_{1a,is}$) remain unchanged from the initial state (potential field).

The outer boundary (with the boundary surface at $x_{1a,ie+1} = 10 R_{\odot}$) is an open outflow boundary, where we simply copy the density, pressure, outflow velocity (inflow set to zero), and the changes of the magnetic fluxes from the outermost layer of cells into the ghost zones as implemented in the following way:
\begin{equation}
\rho (x_{1b,ie+n},x_{2b,j},x_{3b,k}) = \rho (x_{1b,ie},x_{2b,j},x_{3b,k}),n=1,2
\label{eq:rho_iep1}
\end{equation}
\begin{equation}
e (x_{1b,ie+n},x_{2b,j},x_{3b,k}) = e (x_{1b,ie},x_{2b,j},x_{3b,k}),n=1,2
\label{eq:e_iep1}
\end{equation}
\begin{equation}
v_1(x_{1a,ie+n},x_{2b,j},x_{3b,k}) = \mathrm{max} \left[0, \, v_1(x_{1a,ie},x_{2b,j},x_{3b,k}) \right],n=1,2,3
\label{eq:v1_iep1}
\end{equation}
\begin{equation}
v_2(x_{1b,ie+n},x_{2a,j},x_{3b,k}) = v_2(x_{1b,ie},x_{2a,j},x_{3b,k}),n=1,2
\label{eq:v2_iep1}
\end{equation}
\begin{equation}
v_3(x_{1b,ie+n},x_{2b,j},x_{3a,k}) = v_3(x_{1b,ie},x_{2b,j},x_{3a,k}),n=1,2
\label{eq:v3_iep1}
\end{equation}
\begin{equation}
E_1 (x_{1b,ie+n}, x_{2a,j}, x_{3a,k})= E_1 (x_{1b,ie}, x_{2a,j}, x_{3a,k}),n=1,2
\label{eq:E1_iep1}
\end{equation}
\begin{align}
E_2 (x_{1a,ie+n}, x_{2b,j}, x_{3a,k}) & = 2 E_2 (x_{1a,ie+n-1}, x_{2b,j}, x_{3a,k}) g_2 (x_{1a,ie+n-1}) / g_2 (x_{1a,ie+n}) \nonumber \\
& - E_2 (x_{1a,ie+n-2}, x_{2b,j}, x_{3a,k}) g_2 (x_{1a,ie+n-2}) / g_2 (x_{1a,ie+n}),n=2,3
\label{eq:E2_iep3}
\end{align}
\begin{align}
E_3 (x_{1a,ie+n}, x_{2a,j}, x_{3b,k}) & = 2 E_3 (x_{1a,ie+n-1}, x_{2a,j}, x_{3b,k}) g_{31} (x_{1a,ie+n-1}) / g_{31} (x_{1a,ie+n}) \nonumber \\
& - E_3 (x_{1a,ie+n-2}, x_{2a,j}, x_{3b,k}) g_{31} (x_{1a,ie+n-2}) / g_{31} (x_{1a,ie+n}),n=2,3 .
\label{eq:E3_iep3}
\end{align}
In the above, equations (\ref{eq:E1_iep1})-(\ref{eq:E3_iep3}) for specifying the electric fields in the ghost zone cell edges are derived by requiring that the changes of magnetic fluxes through the ghost zone cell faces match those corresponding magnetic flux changes in the outer most layer of cells in the simulation domain.

To initiate the solar wind outflow, we reduce the pressure at the outer boundary from its initial hydrostatic equilibrium value by a factor of 5. The system is then allowed to relax dynamically by numerically solving the MHD equations described in Sections~\ref{sec:equations} and \ref{sec:algorithm}. We run the simulation until a quasi-steady state is achieved, where the initial potential magnetic field is stretched outward into an equatorial streamer.

\begin{figure}[htb!]
\centering
\includegraphics[width=0.8\textwidth]{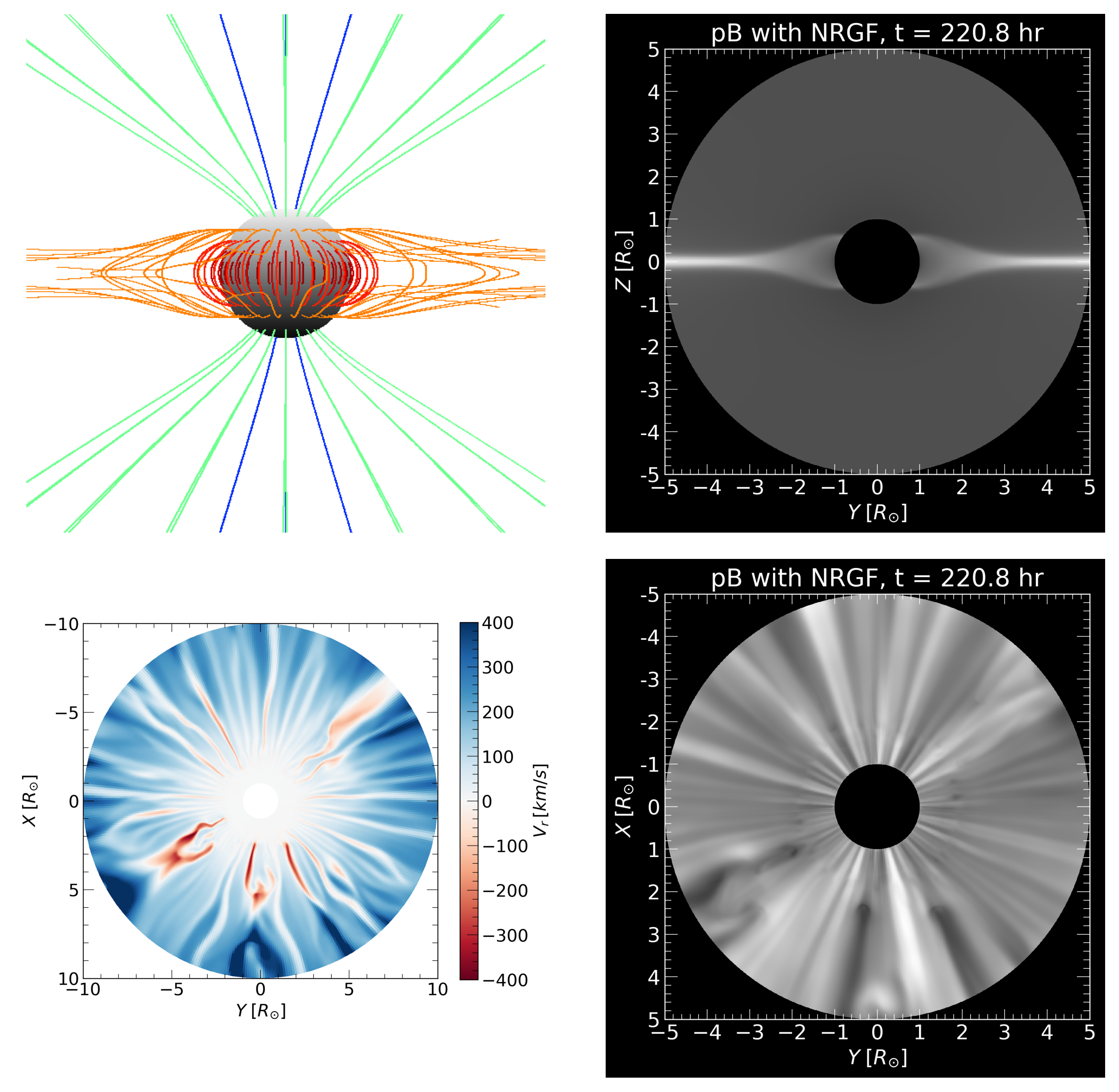}
\caption{Snapshots of the 3D magnetic field lines (top left), radial velocity $v_r$ in the solar equatorial plane (bottom left), and the synthetic white-light coronagraph images (polarized brightness pB with the normalizing radially graded filter (NRGF) by \citet{Morgan:etal:2006} applied) as observed from the equatorial earth view (top right) and from the polar view (bottom right). The gray-scale on the sphere in the top left panel shows the dipolar normal magnetic field distribution $B_r$ on the solar surface. A 22 second animated version of this figure is available in the online journal, which shows about 43 hours of the evolution during the quasi-steady state.}
\label{fig:solwind_snapshots}
\end{figure}

Figure \ref{fig:solwind_snapshots} shows snapshots of the quasi-steady state solution.
In this quasi-steady state, the magnetic field exhibits a partially open configuration (top left panel of Figure~\ref{fig:solwind_snapshots}). The field lines are open (green and blue) at high latitudes and polar regions, while closed field lines (red) dominate the equatorial region beneath cusped field lines (orange), which extend into an equatorial heliospheric current sheet (HCS).  We compute the synthetic white-light coronagraph images (polarized brightness pB \citep[see e.g.][]{Gibson:etal:2016} with the normalizing radially graded filter (NRGF) by \citet{Morgan:etal:2006} applied) as observed from the equatorial earth view (top right panel) and from the polar view (bottom right panel). The synthetic white-light coronagraph image from the Earth-equatorial perspective clearly captures the bright, high-density helmet streamer structures associated with closed and cusped magnetic fields, contrasting with the darker, lower-density coronal holes in the polar regions dominated by open fields. This configuration closely resembles typical solar-minimum corona images observed during eclipses or by coronagraphs. On the other hand, the synthetic polar view highlights dynamic density structures within the heliospheric current sheet, resulting from continuous three-dimensional magnetic reconnection events. These polar-view images reveal elongated, dark (under-dense), wiggly structures flowing sunward, terminating near the top of the helmet streamer. Such features correspond physically to high-speed reconnection jets emanating from active reconnection sites, clearly illustrated by the red sunward velocity streams in the equatorial plane (bottom left panel, Figure~\ref{fig:solwind_snapshots}). These sunward-directed jets exhibit reduced density (appearing darker in white-light) due to flow divergence at the reconnection points. Furthermore, their characteristic wiggly shapes result from the Kelvin-Helmholtz instability driven by strong velocity shear along the jet boundaries. These dynamic structures closely resemble the ``supra-arcade downflows" (SADs) observed in soft X-ray and EUV imaging of solar flare current sheets \citep[e.g.,][]{Savage:etal:2012} and have been successfully reproduced in recent MHD modeling studies \citep[e.g.,][]{Shen:etal:2022}.

\begin{figure}[htb!]
\centering
\includegraphics[width=0.6\textwidth]{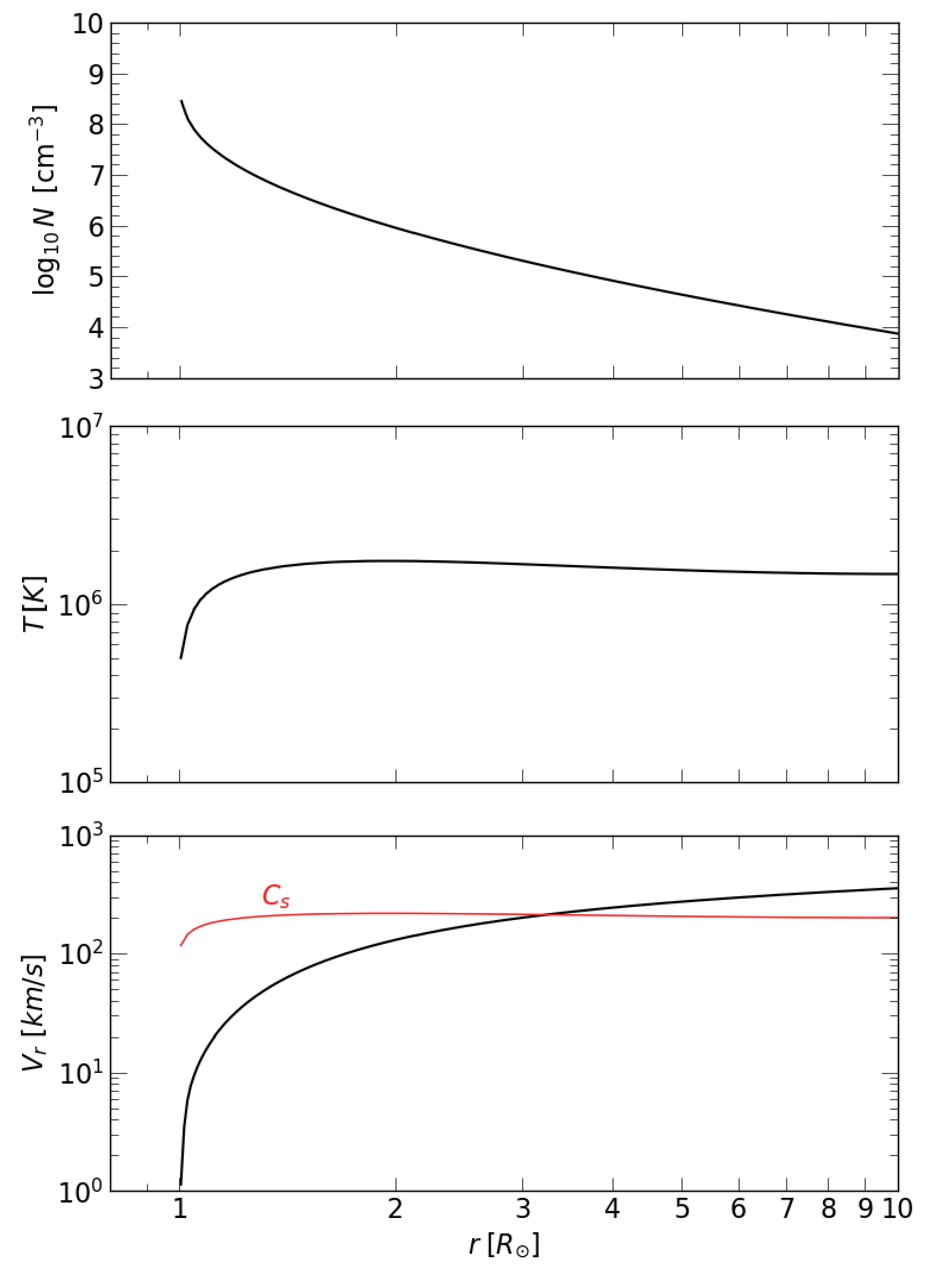}
\caption{The density (upper), temperature (middle), and radial velocity (bottom) profiles along a polar flux tube in the final steady state solar wind solution. The bottom panel also shows the sound speed profile (red curve) along the flux tube. The flow becomes super-sonic at roughly $3 R_{\odot}$, but remains sub-Alfv\'enic throughout the range to $10 R_{\odot}$. Note the Alfv\'en speed is not shown and is mostly above the upper limit of the plot.}
\label{fig:solwind_polartube}
\end{figure}

Within the helmet dome (represented by red field lines in the top-left panel of Figure~\ref{fig:solwind_snapshots}), the closed magnetic field region remains approximately static. This static equilibrium is indicated by the region of negligible radial velocity  $v_r \approx 0$ (white area) extending from $1 \, R_{\odot}$ to approximately $1.8 \, R_{\odot}$ in the equatorial plane (bottom-left panel). Beyond the helmet dome, the open magnetic field region exhibits a steady solar wind outflow. Figure~\ref{fig:solwind_polartube} depicts the steady-state solar wind profiles along a representative polar flux tube.
Our simulated polar flux tube profiles (Figure~\ref{fig:solwind_polartube}) show overall similarity with the 1-D solar-minimum coronal hole model of \citet[][see their Figure 9]{Withbroe:1988}, but exhibit higher density, temperature, and radial velocity values near $10 R_{\odot}$. These differences likely result from complexities inherent in our self-consistent 3D MHD model. For example, the polar flux tube in our simulation undergoes a naturally determined expansion, whereas the 1-D model explicitly imposes a larger, prescribed flux-tube expansion factor. Nevertheless, the flow speed reached at $10 R_{\odot}$, about $357 \, {\rm km/s}$ is within the observed range at that distance (see Figure 9 in \citet{Withbroe:1988}). Thus our test global 3D MHD simulation using a simple dipole magnetic field and a simple empirical coronal heating has produced a reasonable white-light solar corona and the polar solar wind profiles representative of solar minimum configuration. Furthermore our simulation suggests SADs like dynamic features produced by the on-going 3D reconnection in the HCS that can be observed by white-light coronagraph observations from polar viewpoints.

We note that the solar wind outflow becomes super-sonic at about $3 R_{\odot}$ (bottom panel of Figure \ref{fig:solwind_polartube}), yet it remains sub-Alfv\'enic throughout the domain, extending up to $10 R_{\odot}$. The open field regions exhibit very low plasma $\beta$, reaching a minimum value of about 0.004. This result underscores the robustness and numerical stability of our MHD solver under such extreme, magnetically dominated, low plasma-$\beta$ conditions.

We also note that in this solar wind simulation, the inner boundary is placed at the top of the transition region (the base of the corona), so the chromosphere and transition region are not included in the computational domain. The effective response of the chromosphere and transition region to coronal heating is incorporated by adjusting the coronal base pressure to satisfy a balance of the downward heat conduction flux and the integrated radiative cooling across the thin transition region, following \citet{Withbroe:1988}, without explicitly resolving this thin layer. The coronal temperature profile is well resolved in the simulation domain. Consequently, the transition region adaptive conduction (TRAC) method \citep{Johnston:Bradshaw:2019, Johnston:etal:2020, Johnston:etal:2021}, which addresses the numerical difficulty of resolving the transition region, is not necessary for the present simulation. However, for simulations that explicitly resolve the chromosphere and transition region, or that include prominence formation with dynamically evolving prominence-corona-transition regions, implementing TRAC provides a more accurate treatment of coronal density and temperature for quantitative comparison with coronal observables (e.g., EUV emission). Extension of the MFE code to include TRAC for 3D MHD simulations \citep{Johnston:etal:2021} will be pursued in future work.

We have also carried out scaling test runs of the above dipole solar wind simulation which utilizes all the components of the Yin-Yang MFE code and therefore is a good test of the performance of the code.
\begin{figure}[htb!]
\centering
\includegraphics[width=0.6\textwidth]{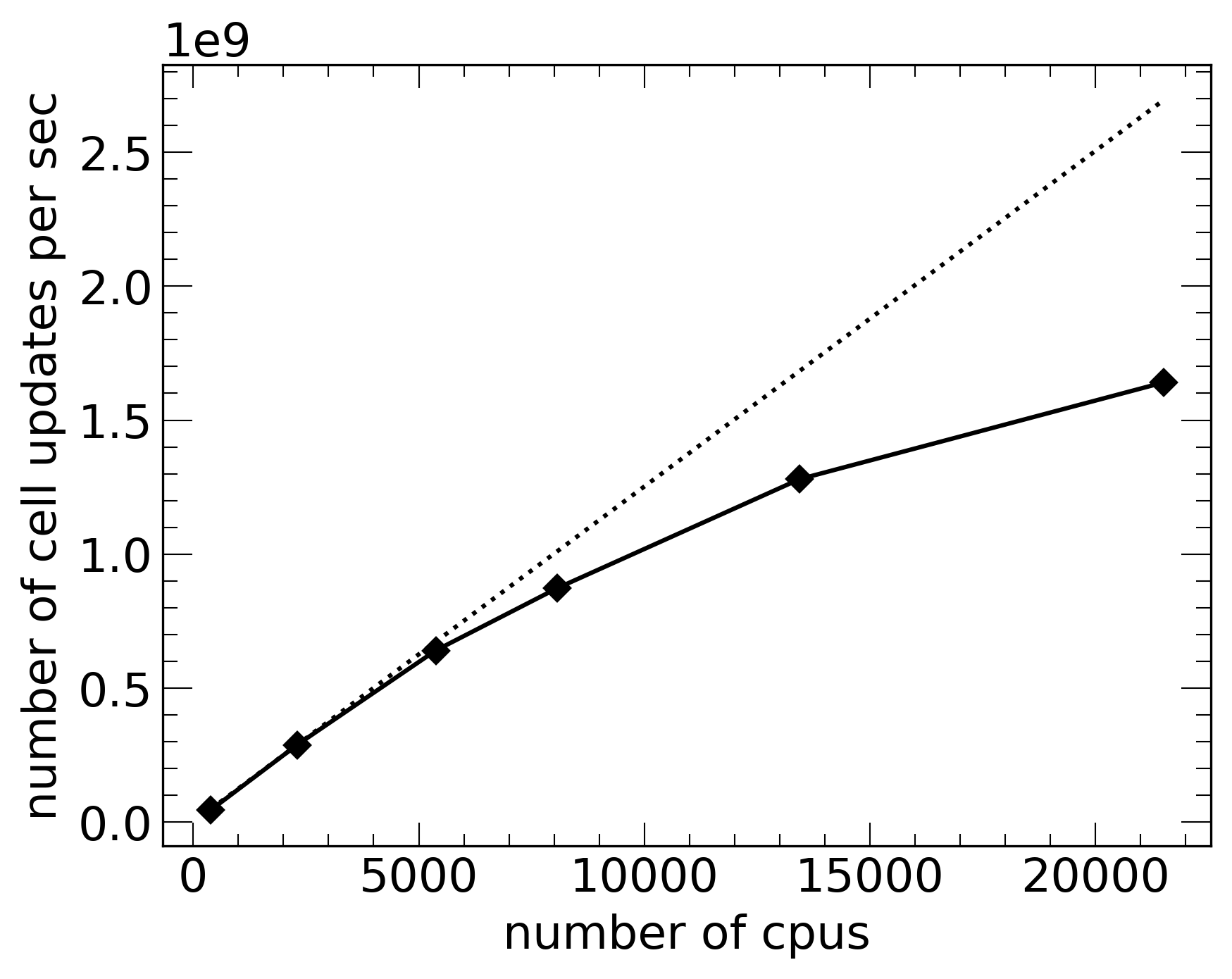}
\caption{Strong scaling of the dipole solar wind simulation using a grid of $840(r) \times 120(\theta) \times 360 (\phi)$ for each Yin and Yang component}
\label{fig:scaling}
\end{figure}
The strong scaling test result (Figure \ref{fig:scaling}) shows good (nearly linear) speed up with the number of CPUs up to about 8000 CPUs for this reasonably large sized Yin-Yang grid with $840(r) \times 120(\theta) \times 360 (\phi)$ for each Yin and Yang component. For this simulation, the code shows a peak performance of about 120,000 cell-updates/cpu/sec. The simulation is run on the HPE Cray EX cluster ``Derecho" of the National Center for Atmospheric Research Wyoming Supercomputing Center (\text{https://ncar-hpc-docs.readthedocs.io/en/latest/compute-systems/derecho/})

\section{Summary}\label{sec:summary}

In this work, we present the Yin-Yang Magnetic Flux Eruption (Yin-Yang MFE) code, a global magnetohydrodynamic simulation tool. The code employs a ZEUS-type staggered grid and updates fluid variables using a modified Lax-Friedrichs scheme applied to the advection terms. A constrained transport method is used to evolve the magnetic field while ensuring it remains divergence-free to machine precision. The scheme supports both second- and fourth-order spatial reconstructions and uses a third-order Runge–Kutta time integrator to advance the MHD equations. To alleviate restrictive time-step constraints in regions with strong magnetic fields and high Alfv\'{e}n speeds, we implement the semi-relativistic (Boris) correction combined with a reduced speed of light. Additional physics modules, including field-aligned thermal conduction (treated using a hyperbolic heat conduction approach), optically thin radiative losses, and empirical coronal heating, are incorporated into the internal energy equation. For full-sphere simulations in spherical coordinates, the Yin-Yang spherical overset grid is utilized to avoid numerical issues and time-step limitations near the poles. 

We validate the numerical methods of the code using a series of standard test problems. Test cases such as the nonlinear polarized Alfv\'{e}n waves, the Brio-Wu shock tube and the Orszag–Tang vortex confirm that the MFE code achieves its designed accuracy and robustly captures nonlinear shocks. We also present full-sphere simulations, including a three-dimensional magnetic field loop advection and a global solar wind model, to demonstrate that the Yin-Yang grid yields stable, accurate solutions and that the physical results are consistent with theoretical expectations. 

Finally, although these numerical schemes form the code’s fundamental framework, Yin-Yang MFE remains under active development, with new physics capabilities and algorithmic improvements continually being integrated. The methods presented here serve as a reference for users interested in understanding the code’s implementation or modifying it for specific applications. Furthermore, the code’s modular design facilitates efficient problem setup and straightforward implementation of new features and modifications.

\acknowledgments
We thank Dr. Binzheng Zhang for helpful discussions throughout this work.
We thank the anonymous referee for helpful comments that improved the paper.
This material is based upon work supported by the National Center for Atmospheric Research (NCAR), which is a major facility sponsored by the U.S. National Science Foundation (NSF) under Cooperative Agreement No. 1852977.
Hongyang Luo is supported by the Research Grants Council (RGC) of Hong Kong General Research Fund (Grant Nos.17309224 and 17308723), and the NSF NCAR HAO's Newkirk graduate fellowship program.
We would like to acknowledge high-performance computing support from the Derecho system (doi:10.5065/qx9a-pg09) provided by the NSF NCAR, sponsored by the NSF.

\end{document}